\documentclass[a4paper,12pt, epsfig]{article}
\pdfoutput=1
\usepackage{epsfig}
\usepackage{epstopdf}
\usepackage{graphicx}
\usepackage{ifthen}

\usepackage{amsmath}
\usepackage[psamsfonts]{amssymb}
\usepackage{euscript}

\usepackage{latexsym}
\usepackage[arrow,matrix,curve]{xy}

\newskip\humongous \humongous=0pt plus 1000pt minus 1000pt

\newif\ifdtup

\def\,{\hspace{-.1cm}}
\def\hsp{,\hspace{.7cm}}

\def\ua{\uparrow}
\def\da{\downarrow}

\def\ms{\mathcal{S}}
\def\tf {\tilde{f}}
\def\fjl {\Phi\left(\frac{n}{q}j,\frac{n}{q} l\right)}
\def\F#1#2{\Phi\left(\frac{n}{q}{#1},\frac{n}{q}{#2}\right)}
\def\fc#1#2 {\frac{n}{q}#1\frac{n}{q}#2}

\def\e#1{\langle#1\rangle_0}
\def\ee#1{\left\langle#1\right\rangle}
\def\di#1{\left(\vcenter{\xymatrix{#1}}\right)}

\def\mf{\mathcal{F}}

\def\s#1{\hbox{\rm sin}\left(#1\right)}
\def\st#1{\hbox{\rm sin}^2\left(#1\right)}
\def\se#1#2{\hbox{\rm sin}^{#1}\left(#2\right)}
\renewcommand{\theequation}{\arabic{section}.\arabic{equation}}
\renewcommand{\(}{\begin{equation}}
\renewcommand{\)}{end{equation} \vspace{-.05in}\linebreak}

\newcounter{saveeqn}
\newcounter{savealpheqn}

\newcommand{\alpheqn}{\setcounter{saveeqn}{\value{equation}}%
  \stepcounter{saveeqn}\setcounter{equation}{0}%
  \renewcommand{\theequation}{\mbox{\arabic{section}.\arabic{saveeqn}
\alph{equation}}}
  \renewcommand{\)}{\end{equation}}}
\def\part#1{\frac{\partial}{\partial{#1}}}%
\def\group#1{\refstepcounter{equation}\setcounter{saveeqn}
 {\value{equation}}%
  \label{#1}\setcounter{equation}{0}%
\renewcommand{\theequation}{\mbox{\arabic{section}.\arabic{saveeqn}
\alph{equation}}}
  \renewcommand{\)}{\end{equation}}}
\newcommand{\reseteqn}{\setcounter{equation}{\value{saveeqn}}%
  \renewcommand{\theequation}{\arabic{section}.\arabic{equation}}%
  \renewcommand{\)}{\end{equation}}}

\newcommand{\aalpheqn}{\setcounter{saveeqn}{\value{equation}}%
  \stepcounter{saveeqn}\setcounter{equation}{0}%
  \renewcommand{\theequation}{\mbox{
        \Alph{subsection}.\arabic{saveeqn}\alph{equation}}}
   \renewcommand{\)}{\end{equation}}}
\newcommand{\areseteqn}{\setcounter{equation}{\value{saveeqn}}%
  \renewcommand{\theequation}{\Alph{subsection}.\arabic{equation}}%
  \renewcommand{\)}{\end{equation}}}

\renewcommand{\thefootnote}{\alph{footnote}}
\renewcommand{\(}{\begin{equation}}
\renewcommand{\)}{\end{equation}}
\newcommand{\ba}{\begin{eqnarray}}
\newcommand{\ea}{\end{eqnarray}}

\newcommand{\bp}{\mathop{\vtop{\ialign{##\crcr
   $\hfil\displaystyle{}\hfil$\crcr\noalign{\kern-13pt\nointerlineskip}
   \BIG{(}\hskip0pt\crcr\noalign{\kern3pt}}}}}
\newcommand{\cbp}{\mathop{\vtop{\ialign{##\crcr
   $\hfil\displaystyle{}\hfil$\crcr\noalign{\kern-13pt\nointerlineskip}
   \BIG{)}\hskip0pt\crcr\noalign{\kern3pt}}}}}
\newcommand{\pa}{\mathop{\vtop{\ialign{##\crcr
    
$\hfil\displaystyle{\oplus}\hfil$\crcr\noalign{\kern+1pt\nointerlineskip 
}
   \hspace{.08in}$^{\alpha=0}$\hskip6pt\crcr\noalign{\kern3pt}}}}}
\renewcommand{\hsp}{,\hspace{.3in}}
\newcommand{\p}{^\prime}

\newcommand{\Z}{\ensuremath{\mathbb Z}}

\catcode`\@=11
\def\vereq#1#2{\lower3pt\vbox{\baselineskip1.5pt \lineskip1.5pt
\ialign{$\m@th#1\hfill##\hfil$\crcr#2\crcr\sim\crcr}}}
\catcode`\@=12

\renewcommand{\(}{\begin{equation}}
\renewcommand{\)}{\end{equation}}

\def\exp#1{\hbox{\rm exp}\left(#1\right)}

\newcommand{\beas}{\begin{eqnarray*}}
\newcommand{\eeas}{\end{eqnarray*}}

\newcommand{\bquo}{\begin{quote}}
\newcommand{\enqu}{\end{quote}}

\newcommand{\C}{{\mathbb C}}
\newcommand{\cp}{{\mathrm{\mathbb CP}}}
\newcommand{\R}{{\mathbb R}}
\renewcommand{\Z}{{\mathbb Z}}

\newcommand{\rhob}{\sigma}

\newcommand{\beq}{\begin{equation}}
\newcommand{\eeq}{\end{equation}}
\newcommand{\bea}{\begin{eqnarray}}
\newcommand{\eea}{\end{eqnarray}}

\newskip\humongous \humongous=0pt plus 1000pt minus 1000pt

\newif\ifdtup

\catcode`\@=11

\@addtoreset{equation}{section}

\def\@normalsize{\@setsize\normalsize{15pt}\xiipt\@xiipt
\abovedisplayskip 14pt plus3pt minus3pt%
\belowdisplayskip \abovedisplayskip
\abovedisplayshortskip \z@ plus3pt%
\belowdisplayshortskip 7pt plus3.5pt minus0pt}

\def\small{\@setsize\small{13.6pt}\xipt\@xipt
\abovedisplayskip 13pt plus3pt minus3pt%
\belowdisplayskip \abovedisplayskip
\abovedisplayshortskip \z@ plus3pt%
\belowdisplayshortskip 7pt plus3.5pt minus0pt
\def\@listi{\parsep 4.5pt plus 2pt minus 1pt
      \itemsep \parsep
      \topsep 9pt plus 3pt minus 3pt}}

\relax

\catcode`@=12

\catcode`\@=11

\def\section{\@startsection{section}{1}{\z@}{3.5ex plus 1ex minus  .2ex}{2.3ex plus .2ex}{\large\bf}}

\def\thesection{\arabic{section}}
\def\thesubsection{\arabic{section}.\arabic{subsection}}

\def\appendix{\setcounter{section}{0}
 \def\thesection{Appendix \Alph{section}}
 \def\thesubsection{\Alph{section}.\arabic{subsection}}
 \def\theequation{\Alph{section}.\arabic{equation}}}
\renewcommand{\theequation}{\arabic{section}.\arabic{equation}}

\begin{document}
\def\thefootnote{\fnsymbol{footnote}}
\def\thetitle{Continuum Limit Matrix Elements for the Tonks-Girardeau Ground State}
\def\autone{Jarah Evslin}
\def\auttwo{Hui Liu}
\def\autthree{Hosam Mohammed}
\def\autfour{Yao Zhou}
\def\affa{Institute of Modern Physics, NanChangLu 509, Lanzhou 730000, China}
\def\affb{University of the Chinese Academy of Sciences, YuQuanLu 19A, Beijing 100049, China}

\begin{center}
{\large {\bf \thetitle}}

\bigskip

\bigskip

{\large \noindent  \autone \footnote{jarah@impcas.ac.cn}, \auttwo \footnote{liuhui@impcas.ac.cn}, \autthree \footnote{hosam@impcas.ac.cn} and \autfour {\footnote{yaozhou@impcas.ac.cn} }}

\vskip.7cm

1) \affa\\
2) \affb\\

\end{center}

\begin{abstract}
\noindent

\noindent
The Tonks-Girardeau model is a quantum mechanical model of $N$ impenetrable bosons in 1+1 dimensions.  A Vandermonde determinant provides the exact $N$-particle wave function of the ground state, or equivalently the matrix elements with respect to position eigenstates.  We consider the large $N$ limit of these matrix elements.  We present a binning prescription which calculates the leading terms of the matrix elements in a time which is independent of $N$, and so is suitable for this limit.  In this sense, it allows one to solve for the ground state of a strongly coupled continuum quantum field theory in the field eigenstate basis.  As examples, we calculate the matrix elements with respect to states with uniform density and also states consisting of two regions with distinct densities.

\end{abstract}

%
\setcounter{footnote}{0}
\renewcommand{\thefootnote}{\arabic{footnote}}




\section{Introduction}

A quantum state is completely characterized by its matrix elements with any basis of the Hilbert space.  These matrix elements generalize the notion of wave function from quantum mechanics to more general quantum systems.  The Coordinate Bethe Ansatz (CBA) \cite{bethe} computes these matrix elements for the coordinate basis, in which the degrees of freedom have concrete values at distinct spatial points.  The CBA only exists for systems with finite numbers of degrees of freedom.  However, these systems all have a number $N$, for example the number of lattice sites in a spin chain or the number of particles in the Tonks-Girardeau \cite{tonks, girardeau} or Lieb-Liniger \cite{ll} models, such that in the formal large $N$ limit one obtains a quantum field theory.  

Were it possible to take the large $N$ limit of the CBA, one would, for the first time, have the explicit wave functions for the states of a strongly interacting quantum field theory.   This would open exciting possibilities.  For example, one could use the equivalence between the high spin XXX spin chain and the $\cp^1$ $\sigma$ model to understand just how the fractional instanton plasma of \cite{grossmeron,ffs} is realized in Minkowski space.  By studying the difference between the ground state and first excited state wave functions, one may achieve an understanding of the nonperturbative generation of a mass gap \cite{affleckmeron} in a full quantum field theory with the same concreteness that is achieved in our understanding of the mass gap in the double well quantum mechanics model.

So far, such a large $N$ limit of general wave functions has been lacking because the Bethe Ansatz is complicated.  It consists of $N!$ terms, each corresponding to an element of the symmetric group $S_N$.  Many approaches have been developed which simplify the calculation of the matrix elements so that it can be done in a time which is only polynomial in $N$.  But this is still too long for the large $N$ limit.  So far, to our knowledge, the large $N$ limit of only one matrix element has ever been computed \cite{brock2}.

In this note we will consider the simplest model which allows a Bethe Ansatz solution, the Tonks-Girardeau model.  In this model we will introduce a novel binning approximation scheme for evaluating the matrix elements in the large $N$ limit.  Our strategy is essentially to decompose $S_N$ into cosets for each of which the sum can be computed explicitly.  For questions of physical interest, like the roles of fractional instantons or unitons in a mass gap, one is interested in matrix elements of Hamiltonian eigenstates with fairly simple coordinate states, or at least with states whose complexities do not themselves depend on $N$.  We will show that for some families of such simple states, our binning prescription allows us to systematically approximate the matrix elements in a time which is independent of $N$.  Therefore we feel that our method can provide a suitable starting point for a large $N$ limit of these states.   

While we suspect that a generalization of our method to any Hamiltonian eigenstate is possible, we restrict our attention to the ground state.  This choice is largely driven by the fact that the ground state matrix elements are all given by Vandermonde determinants.  This allows us to numerically test our approximation scheme for values of $N$ of order $10^4$.    We find that the logarithms of the matrix elements can be expanded in powers of $1/N$ and our method allows the leading term in this expansion to be estimated.  In some cases we attempt to improve the accuracy of this estimation and we find that we can increase the accuracy as desired at the cost of a longer computation.  Critically, the length of computation never depends on $N$, but only on the complexity of the state and the desired precision.

We begin in Sec.~\ref{tgsez} with a review of the Tonks-Girardeau model and its solution using the Bethe Ansatz.  Next in Sec.~\ref{consez} we consider matrix elements with states in which the gas has a constant density.  First we give an explicit and exact formula for the matrix elements using Vandermonde determinants.  Next we introduce our binning approach and use it to calculate a set of approximations of the matrix elements of ever increasing accuracy.  Finally, in Sec.~\ref{doubsez}, we consider matrix elements with respect to states with two densities.  We consider two cases, one in which one density vanishes and another in which the region with each density contains the same number of particles.  As this case is more complicated than the single density case, we compute only the leading approximation.   We find that it calculates the logarithm of the matrix elements with an error which is roughly independent of $N$, and in general less than 10\%.

\section{The Tonks-Girardeau Model} \label{tgsez}

\subsection{The Model and Its Ground State}

A Tonks-Girardeau gas  is a (1+1)-dimensional, nonrelativistic quantum mechanical model of $N$ impenetrable bosons on a circle of circumference $L$.  The wave function of the $N$ particles satisfies the free nonrelativistic Schrodinger equation
\beq
-\frac{\hbar^2}{2m}\frac{\partial^2\psi(\bf{x})}{\partial x_j^2}=E\psi
\eeq
where
\beq
{\mathbf{x}}=\{x_j\}\hsp j\in[1,N]\hsp x_j\in[0,L]
\eeq
are the coordinates of the $N$ particles.  The impenetrability comes from the additional condition
\beq
\psi({\mathbf{x}})=0{\textrm{\ if\ }} x_j=x_l \label{imp}
\eeq
for any $j,\ l$.  Periodic boundary conditions are assumed
\beq
\psi({\bf{x\p}})=\psi({\bf{x}}){\textrm{\ if\ }} x\p_j=x_j{\textrm{\ mod\ }} L \label{per}
\eeq
for all $j$.

The periodicity condition (\ref{per}) restricts the space of wave functions to the subspace of free particle states generated by $e^{2 \pi i k x/L}$ with $k\in\Z$.  The impenetrability condition (\ref{imp}) further restricts the space of wave functions to the subspace of Slater determinants of $N$ such plane waves
\beq
e^{2\pi i k_i x/L}\hsp i\in [1,N]\hsp k_i\in \Z.
\eeq
In other words, if one defines the matrix
\beq
M_{ij}({\bf{x}})=e^{2\pi i k_i x_j/L}
\eeq
then
\beq
\psi({\bf x})={\rm{Det}}\left(M_{ij}({\bf x})\right). \label{wf}
\eeq
Note that if two $k_i$ are equal then the wave function $\psi({\bf x})$ vanishes.

The ground state wave function can and will be taken to be real and positive.    It corresponds to the case
\beq
k_i=i-\frac{N+1}{2}.
\eeq
We will impose the positivity condition by simply taking the absolute value of $\psi$ in each expression
\beq
\psi_0({\bf x})=\left|{\rm{Det}}\left(M_{ij}({\bf x})\right)\right|. \label{gs}
\eeq
Adding a constant to the $k_i$ has the effect of multiplying the wave function by an $x$-dependent scalar which would change the state (\ref{wf}) and in general even change the energy.   However, the expression (\ref{gs}) for the ground state wave function, unlike (\ref{wf}), is invariant under this operation because the overall phase is zeroed by the absolute value.  Therefore, for simplicity, we will add $(N-1)/2$ to $k_i$ and so set
\beq
k_i=i-1.
\eeq
This still describes the ground state wave function so long as we use the expression (\ref{gs}).  Note that the $k_i$ can only be interpreted as momenta if the wave function is defined by Eq.~(\ref{wf}) and so we have not actually modified the momenta, or even the state.

With this choice of $k_i$, the wave function can be brought into Vandermonde form.  Defining the Vandermonde matrix
\beq
\tilde{M}_{ij}=\eta_j^{(i-1)}\hsp \eta_j=e^{2\pi i x_j/L} \label{vanmat}
\eeq
the wave function is a Vandermonde determinant
\beq
\psi_0({\bf x})=\left|{\rm{Det}}\left(\tilde{M}_{ij}({\bf x})\right)\right|
=\prod_{i=1}^{N-1}\prod_{j=i+1}^N\left|\eta_i-\eta_j\right|. \label{gsvan}
\eeq

\subsection{A Bethe Ansatz Formulation}
If the $x_j$ or $k_i$ are evenly spaced then the wave function can be brought into Vandermonde form.  Whether or not this is the case, the determinant formulas above can all be expanded using the Leibnitz formula for the determinant.  We will now provide the expansion of ({\ref{gsvan}), the following proceeds identically for (\ref{wf}) and (\ref{gs})

\beq
\psi_0({\bf x})=\left|\sum_{g\in S_N}(-1)^g\prod_{j=1}^N \tilde{M}_{j P(j)}\right|
=\left|\sum_{g\in S_N}(-1)^g \exp{\sum_{j=1}^N \frac{2\pi i x_j P(j)}{L}}\right|.
\eeq
Here $g$ is a permutation of the set of natural numbers $[1,N]$, realized by the bijection $P:[1,N]\mapsto[1,N]$.  The symbol $(-1)^g$ is equal to $1$ for an even permutation and $(-1)$ for an odd permutation.

Let us define 
\beq
m_j=x_j \frac{N}{L}\hsp
K_j=2\pi \frac{j}{N}\hsp
\Psi_{jk}=2\pi \theta(j-k) \label{bethevar}
\eeq
where $\theta$ is the Heaviside step function.  Then
\beq
\psi_0({\bf x})=\left|\sum_{g\in S_N}\exp{i\sum_{j=1}^N m(j)K_{P(j)}+\frac{i}{2}\sum_{j<k}^N\Psi_{P(j),P(k)}}\right|. \label{bethe}
\eeq
This is the form of the Bethe Ansatz \cite{bethe}, which also describes many more interesting systems.  The goal of this paper is to introduce a new method to perform the large $N$ limit of the sum (\ref{bethe}).  The fact that this sum admits a Vandermonde form (\ref{gsvan}) will allow numerical checks of our formulas.

\section{A Constant Density Gas} \label{consez}

We will begin with the simplest case
\beq
x_j=j\frac{L}{N} \label{cons}
\eeq
corresponding to a constant density gas.  In Subsec.~\ref{essez} we will use the Vandermonde form (\ref{gsvan}) of the ground state wave function to evaluate the ground state wave function $\psi_0({\bf x})$ at this point, or equivalently to find the matrix element
\beq
\langle {\bf x} | \Omega\rangle
\eeq
between the ground state wave function $|\Omega\rangle$ and the state $|{\bf x}\rangle$ corresponding to a constant density gas (\ref{cons}).  

Next we will calculate the same matrix element using our binning formalism.  We will find that the binning formalism estimates the logarithm of the answer.  Successive approximations will be made which calculate this logarithm more and more precisely.  The complexity of this calculation is independent of $N$.

\subsection{The Exact Solution} \label{essez}

Let us define the $N$th root of unity
\beq
\chi=e^{2\pi i/N}.
\eeq
Then the Vandermonde form of the ground state wave function (\ref{gsvan}) may be simplified as follows
\bea
\psi_0({\bf x})&=&\prod_{i=1}^{N-1}\prod_{j=i+1}^N\left|\chi^i-\chi^j\right|=\prod_{i=1}^{N-1}\prod_{j=i+1}^N\left| 1-\chi^{(j-i)}\right|\label{vform}\\
&=&\prod_{k=1}^{N-1}\left| 1-\chi^k\right|^{N-k}
=\prod_{k=1}^{N-1}\left| 1-\chi^k\right|^{N/2}
=\left|\prod_{k=1}^{N-1} \left(1-\chi^k\right)\right|^{N/2}.
\nonumber
\eea

Therefore we need to calculate the term $\beta_N$ in the absolute value
\beq
\psi_0({\bf x})=\left|\beta_N\right|^{N/2}\hsp \beta_N=\prod_{k=1}^{N-1} \left(1-\chi^k\right).
\eeq
Here $\beta_N$ is an $(N-1)$st order polynomial in the $\chi$.  Let $\gamma_{N,j}$ be the term with $j$ powers of $\chi$
\beq
\beta_N=\sum_{j=0}^{N-1} (-1)^j \gamma_{N,j}.
\eeq

We will proceed by calculating all of the $\gamma_{N,j}$.  As a warm up, let us start with small values of $j$.  Clearly 
\beq
\gamma_{N,0}=1\hsp \gamma_{N,1}=\sum_{i=1}^{N-1}\chi^i=-1.
\eeq
What about $j=2$?
\beq
\gamma_{N,2}=\sum_{i_1=1}^{N-2}\sum_{i_2=i_1+1}^{N-1} \chi^{i_1+i_2}=\sum_{i=3}^{2N-3} a^{(2)}_i \chi^i
\eeq
where $a^{(2)}_i$ is the number of pairs ${i_1,i_2}$ with $1\leq i_1<i_2\leq N-1$ and $i=i_1+i_2$.  Let us defined the reduced summands
\beq
\tilde{i}_j=i-j\hsp 0\leq \tilde{i}_1\leq \tilde{i}_2\leq N-3 \hsp i-3=\tilde{i}_1+\tilde{i}_2.
\eeq
Now the set $\{\tilde{i}_k\}$ is a two element partition of $i-3$.  It is conventional to count the number of nonzero elements in a partition, so then this is a partition of at most two elements.  The constants $a^{(2)}_i$ count those partitions of $i-3$ of at most two elements in which each element is bounded by $N-3$.  This number is conventionally denoted
\beq
a^{(2)}_i=p(2,N-3;i-3).
\eeq
These functions are known to generate the Gaussian binomial coefficients
\beq
\sum_{i=0}^{MN} p(M,N;i)\chi^i={{M+N}\choose{M}}_\chi=\frac{\prod_{j=N+1}^{M+N}\left(1-\chi^j\right)}{\prod_{j=1}^{M}\left(1-\chi^j\right)}. \label{wiki}
\eeq
Putting this all together
\bea
\gamma_{N,2}&=&\sum_{i=3}^{2N-3}p(2,N-3;i-3)\chi^i=\chi^3\sum_{i=0}^{2(N-3)}p(2,N-3;i)\chi^{i}=\chi^3 {{N-1}\choose 2}_\chi\nonumber\\
&=&\chi^3\frac{\prod_{j=N-2}^{N-1}\left(1-\chi^j\right)}{\prod_{j=1}^{2}\left(1-\chi^j\right)}=\chi^3\frac{(1-\chi^{N-1})(1-\chi^{N-2})}{(1-\chi)(1-\chi^2)}\nonumber\\
&=&\chi^3\frac{(1-\chi^{-1})(1-\chi^{-2})}{(1-\chi)(1-\chi^2)}=(-1)^2=1
\eea
where we used the fact that $\chi^N=1$.

The general case $\gamma_{N,M}$ 
\beq
\gamma_{N,M}=\sum_{\{i_j\}}(-1)^{N-1} \chi^{\sum_{j=1}^M i_j }\hsp 1\leq i_1<i_2 ... i_M\leq N-1
\eeq
is no more difficult.   Again one may define the coefficients $a^{(M)}_i$ by
\beq
\gamma_{N,M}=\sum_{i=M(M+1)/2}^{MN-M(M+1)/2} a^{(M)}_i \chi^i.
\eeq
These count the allowed values of $\{i_j\}$ or equivalently the reduced
\beq
\tilde{i}_j=i_j-j\hsp 0\leq \tilde{i}_1\leq \tilde{i}_2 ... \tilde{i}_M\leq N-M-1\hsp i-\frac{M(M+1)}{2}=\sum_{j=1}^M\tilde{i}_j.
\eeq
These are partitions of $i-M(M+1)/2$ consisting of at most $M$ elements such that the maximal value is $N-M-1$ so
\beq
a^{(M)}_i =p\left(M,N-M-1;i-\frac{M(M+1)}{2}\right).
\eeq
Finally we may use (\ref{wiki}) to evaluate $\gamma_{N,M}$
\bea
\gamma_{N,M}&=&\sum_{i=M(M+1)/2}^{MN-M(M+1)/2} p\left(M,N-M-1;i-\frac{M(M+1)}{2}\right)\chi^i\nonumber\\
&=&\chi^\frac{M(M+1)}{2}\sum_{i=0}^{2\left(N-\frac{M(M+1)}{2}\right)}p(M,N-M-1;i)\chi^{i}=\chi^\frac{M(M+1)}{2}\ {{N-1}\choose M}_\chi\nonumber\\
&=&\chi^\frac{M(M+1)}{2}\frac{\prod_{j=N-M}^{N-1}\left(1-\chi^j\right)}{\prod_{j=1}^{M}\left(1-\chi^j\right)}=\frac{\prod_{j=1}^{M}\left(1-\chi^{-j}\right)\chi^j}{\prod_{j=1}^{M}\left(1-\chi^j\right)}=(-1)^M.
\eea
We then find
\beq
\beta_N=\sum_{j=0}^{N-1} (-1)^j \gamma_{N,j}=\sum_{j=0}^{N-1} 1=N
\eeq
and so the matrix element is
\beq
\psi_0({\bf{x}})=N^{N/2}. \label{esatta}
\eeq

It may seem odd that the matrix element grows with $N$, but using our conventions
\beq
\langle \Omega|\Omega\rangle=\int d{\bf x}|\psi({\bf x})|^2=N!L^N
\eeq
and the norm of the $|{\bf x}\rangle$ states is infinite.  To obtain probabilities from these matrix elements, of course one would need to normalize the states.  This would not be difficult, but our goal is just to calculate the matrix elements.

\subsection{The Binning Procedure} \label{binsez}

The dramatic simplifications in the above calculation occur in part because in Eq.~(\ref{bethevar}) $m_j$ and $K_j$ are linear while $\Psi_{jk}$ is quite simple.  The simplicity of $\Psi_{jk}$ occurs for all matrix elements in the model, and that of $K_j$ occurs for all matrix elements of the ground state $|\Omega\rangle$.  However, to evaluate the ground state wave function with general particle positions ${\bf x}$ one needs to relax the linearity of $m_j$ or equivalently of $x_j$.  

We are interested in the limit of a large number $N$ of particles.  In this limit, it makes sense to consider a local density of particles.  We do not know whether the wave function will be supported on configurations for which the local density is smooth.  Nonetheless, the matrix elements with respect to smooth configurations are the most interesting for physical applications.  Our goal is to devise a method to compute such matrix elements.  

Our wave function is symmetric under permutations of the $x_j$ \cite{girardeau} and so without loss of generality we may order the $x_j$ so that $x_{j+1}>x_j$.  Then if the density is smooth, on sufficiently small scales the $x_j$ will be linearly distributed in $j$ to any desired precision for $j$ in a sufficiently small interval.   Therefore we expect simplifications to occur on such small scales.  Our strategy will be to decompose the wave function into factors which depend only on such small ranges, to evaluate these factors and then to reassemble them to obtain an approximate wave function.

More concretely, our procedure for evaluating the sum (\ref{bethe}) is as follows:

\noindent
{\bf{Step 1:}} Divide the ordered set $\{x_j\}$ into $q$ intervals $\ms_i$ called bins.

\noindent
{\bf{Step 2:}} For each bin $\ms_i$ with $n_i$ elements, choose a disjoint subset $F_i\subset[1,N]$ which also has $n_i$ elements.   Recall that each $g\in S_N$ corresponds to a map $P:[1,N]\longrightarrow [1,N]$.  Let $H\subset G$ consist of all elements $g\in S_N$ such that the corresponding map satisfies
\beq
P(\ms_i)=F_i.
\eeq

\noindent
{\bf{Step 3:}} Calculate the terms in Eq.~(\ref{bethe}) corresponding to $H\subset S_N$.  In other words, sum over only those permutations which fix $P(\ms_i)=F_i$.


\noindent
{\bf{Step 4:}} Consider all possible sets of images $\{F_i\}$ and sum over the result of step 3 for each summand.

This procedure, if carried out exactly, will produce the exact wave function.  However step 4 is computationally difficult and so we will only perform this sum approximately, yielding an approximate answer.

\subsection{The Two Bin Approximation: The Master Formula} \label{binmsez}
Let $N$ be even.  We will divide $[1,N]$ into two bins
\beq
\ms_A=[1,N/2]\hsp \ms_B=[N/2+1,N]. \label{msd}
\eeq
Choose a subset $F\subset[1,N]$ consisting of $N/2$ elements.  Recall that each permutation $g\in S_N$ corresponds to a function $P:[1,N]\longrightarrow [1,N]$.  Let $H_F\subset S_N$ be the subgroup such that if $g\in H_F$ then the corresponding function $P$ satisfies
\beq
P(\ms_A)=F. \label{fdef}
\eeq
We will refer to $F$ as the image, reflecting Eq.~(\ref{fdef}).

We have thus completed steps 1 and 2.  The $q=2$ intervals are $\ms_A$ and $\ms_B$ and the image $P(\ms_A)$ is $F$, whereas $P(\ms_B)$ is the complement of $F$ in $[1,N]$.  Next for step 3.  Recall that we are still considering the matrix element of the ground state and the constant density state.  We need to evaluate
\bea
\psi_F({\bf x})&=&\sum_{g\in H_F}\exp{i\sum_{j=1}^N m(j)K_{P(j)}+\frac{i}{2}\sum_{j<k}^N\Psi_{P(j),P(k)}}\nonumber\\
&=& \sum_{g\in H_F} (-1)^g \chi^{\sum_{j=1}^N j P(j)}. \label{passo4}
\eea
To evaluate this, we need to understand $H_F$.  

Elements of $H_F$ are represented by maps $P$ such that $P(\ms_A)=F$.  For a given image $F$, choose one such map $P_F$ corresponding to any fixed group element $g_F\in H_F$.  Any other element of $H_F$ can be obtained by acting on $g_F$ from the right by a permutation of the image $F$ and of its complement $[1,N]\setminus F$.  These two permutation groups are each $S_{N/2}$ and so $H_F$ is a right coset
\beq
H_F=g_F (S_{N/2}\times S_{N/2}).
\eeq
Similarly, any element of $H_F$ can be obtained by acting on $g_F$ from the left by a permutation $S_{N/2}$ of $\ms_A$ and a permutation $S_{N/2}$ of $\ms_B$ and so $H_F$ is also a left coset
\beq
H_F=(S_{N/2}\times S_{N/2})g_F.
\eeq
Using either representation, the key point is that the two copies of $S_{N/2}$ commute with each other and so (\ref{passo4}) can be factored into one expression for each copy
\bea
\psi_F({\bf x})&=&(-1)^{g_F}\psi_A \psi_B\\
\psi_A&=& \sum_{g_A\in S_{N/2}}(-1)^{g_A}\chi^{\sum_{j=1}^{N/2}jP_A(j)}\hsp \label{fact}
\psi_B= \sum_{g_B\in S_{N/2}}(-1)^{g_B}\chi^{\sum_{j=N/2+1}^{N}jP_B(j)}\nonumber
\eea
where
\beq
P_A:\ms_A\longrightarrow F\hsp
P_B:\ms_B\longrightarrow [1,N]\setminus F
\eeq
represent $g_Ag_F$ and $g_Bg_F$ respectively.


\noindent
{\bf{Simplification 1:}} Up to a total phase, $\psi_A$ and $\psi_B$ are Vandermonde determinants.    This is true not only for the ground state, but also for all excited states.  Each $\psi_I$ is always a Slater determinant and it is equal to a phase times the Vandermonde determinant whenever the $x_j$ with $j$ in the bin $j\in\ms_I$ are evenly spaced.

In more general models whose states can be put in Bethe Ansatz form, matrix elements are not given by Vandermonde determinants.  However, simple cases in which the $x$ are distributed linearly, such as the N\'eel state in the XXX spin chain, do have known matrix elements \cite{llmat,brock1,brock2}.  Therefore we hope that in general our binning method will reduce the calculation of piecewise linear $x$ matrix elements to that of linear $x$ matrix elements, which at least in some cases can be found via traditional methods.

To bring $\psi_A$ into Vandermonde form, we pull out an overall phase
\beq
\psi_A=\chi^{\sum_{j\in F} j}\sum_{g_A\in S_{N/2}}(-1)^{g_A}\chi^{\sum_{j=0}^{N/2-1}jP_A(j+1)}
=\chi^{\sum_{j\in F} j}\prod_{j<k}^{N/2}\left(\chi^{P_A^F(k)}-\chi^{P_A^F(j)}\right) \label{avan}
\eeq
where $P_A^F$ is the map $P_A:[1,N/2]\rightarrow F$ obtained from the action of $g_F$ on $[1,N/2]$.  Note that each $\chi^j$ appears in the product $N/2-1$ times.  And so we may pull out the midpoints of the exponents from each factor to obtain
\bea
\psi_A&=&\chi^{\left(\frac{N/2-1}{2}+1\right)\sum_{j\in F} j}\prod_{j<k}^{N/2}\left(\chi^{\frac{1}{2}\left(P_A^F(k)-P_A^F(j)\right)}-\chi^{\frac{1}{2}\left(P_A^F(j)-P_A^F(k)\right)}\right)\nonumber\\
&=&(2i)^{\frac{N}{4}\left(\frac{N}{2}-1\right)}\chi^{\left(\frac{N}{4}+\frac{1}{2}\right)\sum_{j\in F} j}\prod_{j<k}^{N/2}{\rm{sin}}\left(\frac{\pi}{N}\left(P_A^F(k)-P_A^F(j)\right)\right). \label{sin}
\eea
Recall that we are free to choose any $g_F$ such that $P_A^F(\ms_A)=F$.  Let us fix $g_F$ by demanding that the $P_A^F(j)$ are monotonically increasing with $j$.  In this case, the arguments $\frac{\pi}{N}\left(P_A^F(k)-P_A^F(j)\right)$ of the sine functions on the right hand side of (\ref{sin}) are strictly positive.  

The factor $\psi_B$ can similarly be put in Vandermonde form
\bea
\psi_B&=&\chi^{\left(\frac{N}{2}+1\right)\sum_{j\in [1,N]\setminus F} j} \sum_{g_B\in S_{N/2}}(-1)^{g_B}\chi^{\sum_{j=0}^{N/2-1}jP_B(j+N/2+1)}\\
&=&(-1)^{\sum_{j\in F} j}\chi^{\sum_{j\in [1,N]\setminus F} j}\sum_{g_B\in S_{N/2}}(-1)^{g_B}\chi^{\sum_{j=0}^{N/2-1}jP_B(j+N/2+1)}.\nonumber
\eea
This is identical to the expression (\ref{avan}) for $\psi_A$ except for an overall $F$-dependent sign.  Therefore
\beq
\psi_B=(-1)^{\sum_{j\in F} j}(2i)^{\frac{N}{4}\left(\frac{N}{2}-1\right)}\chi^{\left(\frac{N}{4}+\frac{1}{2}\right)\sum_{j\in [1,N]\setminus F} j}\prod_{j<k}^{N/2}{\rm{sin}}\left(\frac{\pi}{N}\left(P_B^F(k+N/2)-P_B^F(j+N/2)\right)\right).
\eeq
Again, we will choose $g_F$ such that $P_B^F$ is monotonically increasing, so the arguments of the sine functions will be positive.

Multiplying $\psi_A$ by $\psi_B$, the $F$-dependence drops out of the $\chi$ term, as the $j$ sum is now over the entire range $[1,N]$.  Therefore the $F$-dependence of the phase of $\psi_F$ lies only in the overall signs $(-1)^{g_F}(-1)^{\sum_{j\in F} j}$
\bea
\psi_F({\bf x})&=&(-1)^{g_F}(-1)^{\sum_{j\in F} j}(2i)^{\frac{N}{2}\left(\frac{N}{2}-1\right)}\label{ff}\\
&&\times\prod_{j<k}^{N/2}{\rm{sin}}\left(\frac{\pi}{N}\left(P_A^F(k)-P_A^F(j)\right)\right){\rm{sin}}\left(\frac{\pi}{N}\left(P_B^F(k+N/2)-P_B^F(j+N/2)\right)\right).\nonumber
\eea

Now we will show that this phase is just
\beq
(-1)^{g_F}(-1)^{\sum_{j\in F} j}=(-1)^{\frac{N}{4}\left(\frac{N}{2}+1\right)} \label{fasa}
\eeq
and so it is also independent of the choice of image $F$.  First, consider the case $F=[1,N/2]$.  In this case $g_F$ is the identity permutation.  Therefore
\beq
(-1)^{g_F}=1\hsp (-1)^{\sum_{j\in F} j}=(-1)^{\sum_{j=1}^{N/2}j}=(-1)^{\frac{N}{4}\left(\frac{N}{2}+1\right)} \label{id}
\eeq
and so (\ref{fasa}) is satisfied.  

Next consider a single permutation which exchanges two elements $j,\ k\in[1,N]$.  Let $j<k$.  If both $j$ and $k$ are in $F$ or in $[1,N]\setminus F$ then $F$ is invariant and so the quantity (\ref{fasa}) cannot change.  For concreteness, let $j\in F$ and $k\in[1,N]\setminus F$.  Then a permutation which exchanges $j$ and $k$, acting on $g$ from the right, will yield a new image
\beq
F\p=\left(F\setminus\{j\}\right)\cup\{k\}.
\eeq
Let $j\p$ be the number of elements of $F$ which are less than $j$ and let $k\p$ be the number of elements of $F$ which are less than $k$.  Since there are $j-1$ and $k-1$ elements which are lower than $j$ and $k$ in $[1,N]$, then there must be $j-j\p-1$ elements of $[1,N]\setminus F$ which are lower than $j$ and $k-k\p-1$ which are lower than $j$.  

Now we are ready to calculate the change in $(-1)^{g_F\p}$ resulting from the exchange of $j$ and $k$.  Note that the resulting permutation $g_{F\p}$ is not just the transposition $(jk)$.  This is because once $j$ and $k$ have been exchanged, the elements of $F\p$ and $[1,N]\setminus F\p$ are no longer in order.  $k$ now appears in $F$ in the old position of $j$, which has $j\p$ elements before it.  However $k$ has $k\p$ elements beneath it in $F$.  One of these elements is $j$, which is not present in $F\p$, and so $k$ has $k\p-1$ elements beneath it in $F$.  The result is that in order to order $F\p$, $k$ needs to be moved to the right $k\p-j\p-1$ places, requiring $k\p-j\p-1$ transpositions.  Similarly, $j$ now appears after $k-k\p-1$ elements of $[1,N]\setminus F\p$ but it is only greater than $j-j\p-1$ elements, and so to order $[1,N]\setminus F\p$ requires $(k-j-k\p+j\p)$ transpositions.  Summing the transposition $(jk)$ and the transpositions in $F\p$ and its complement we find
\beq
(-1)^{g_F\p}(-1)^{g_F}= (-1)^{1+(k\p-j\p-1)+(k-j-k\p-j\p)}=(-1)^{k-j}.
\eeq
The change in the other sign factor is
\beq
(-1)^{\sum_{j\in F} j}(-1)^{\sum_{j\in F\p} j}=(-1)^{2\sum_{j\in F\cap F\p} j}(-1)^{j+k}=(-1)^{j+k}=(-1)^{g_F\p}(-1)^{g_F}
\eeq
and so
\beq
(-1)^{g_F}(-1)^{\sum_{j\in F} j}=(-1)^{g_F\p}(-1)^{\sum_{j\in F\p} j}.
\eeq
Therefore the phase factor is invariant under all interchanges of two elements.  But one may obtain any image $F\p$ via interchanges of pairs of elements, and so the phase factor is the same for all $F$, therefore it is always equal to its value when $F=[1,N/2]$ which is given in Eq.~(\ref{id}).  Therefore we have proven (\ref{fasa}).

Substituting (\ref{fasa}) into (\ref{ff}) we find our master formula for the contribution of one choice of image $F$ to $\psi_0$
\beq
\psi_F({\bf x})=(-1)^{\frac{N}{4}\left(\frac{N}{2}+1\right)} (2i)^{\frac{N}{2}\left(\frac{N}{2}-1\right)} \prod_{j<k}^{N/2}{\rm{sin}}\left(\frac{\pi}{N}\left(P_A^F(k)-P_A^F(j)\right)\right){\rm{sin}}\left(\frac{\pi}{N}\left(P_B^F(k+N/2)-P_B^F(j+N/2)\right)\right) \label{master}
\eeq
completing step 3.

Recall that the sine terms are all positive and the phases are manifestly independent of $F$, and so we have arrived at

\noindent
{\bf{Simplification 2:}} The phase of $\psi_F$ is independent of $F$.

\subsection{An All Even Image} \label{parisez}
As we may decompose the symmetric group into cosets
\beq
S_N=\coprod_F H_F
\eeq
we may also decompose the matrix elements into terms of the form (\ref{master})
\beq
\psi_0=\left|\sum_F \psi_F\right|.
\eeq
This is step 4 of our method.

As the contributions $\psi_F$ to $\psi_0$ are all of the same phase, our approach will be to find the largest and then to expand about it.  The sine functions are maximized when the distances between the elements of $F$ and of its complement are largest.  Therefore a reasonable guess at the largest $\psi_F$ is the case when $F$ consists of all even or all odd elements of $[1,N]$.  

Let us introduce some notation.  Define the vector
\beq
\mf_i=\left\{\begin{tabular}{ll}$1$ \ \ if $i\in F$\\$0$\ \ \   otherwise\\
\end{tabular} \right. \hsp
i\in [1,N].
\eeq
So $\mf$ is a sequence of $N$ $0s$ and $1s$ with $N/2$ of each.  Let us name two common subsequences
\beq
\ua=\{1,0\}\hsp\da=\{0,1\}. \label{frdef}
\eeq
So the all even subset $F$ corresponds to $\mf=\da\cdots\da$ with $N/2$ $\da$'s.   Similarly the odd subset $F$ corresponds to $N/2$ $\ua$'s.  For example, if $\mf=\{0,1,1,0\}$ then one could write the shorthand $\mf=\da\ua$.

In Eq.~(\ref{master}) one sees that $\psi_F$ only depends upon the differences between the $P_A^F(j)$.  These differences are identical for the even and odd subsets, so they yield the same $\psi_F$.  We will now calculate $\psi_F$ for the even subset
\beq
P_A^{\da\cdots\da}(j)=2j\hsp P_B^{\da\cdots\da}(j+N/2)=2j-1\hsp
j\in\ms_A=[1,N/2].
\eeq
Now (\ref{master}) becomes
\beq
\psi_{\da\cdots\da}({\bf x})=(-1)^{\frac{N}{4}\left(\frac{N}{2}+1\right)}
 (2i)^{\frac{N}{2}\left(\frac{N}{2}-1\right)} \prod_{j<k}^{N/2}{\rm{sin}}^2\left(\frac{2\pi}{N}(k-j)\right). \label{xc}
\eeq
The term on the right looks quite similar to the first expression in (\ref{vform}).

In that case we saw
\beq
\psi_0=\prod_{j<k}^N\left|\chi^j-\chi^k\right|=\prod_{j<k}^N\left|\chi^{(j-k)/2}-\chi^{(k-j)/2}\right|=
(2i)^{N(N-1)/2}\prod_{j<k}^N{\rm{sin}}\left(\frac{\pi}{N}(k-j)\right)=N^{N/2}.
\eeq
Replacing $N$ by $N/2$ one finds
\beq
(2i)^{\frac{N}{4}\left(\frac{N}{2}-1\right)}\prod_{j<k}^{N/2}{\rm{sin}}\left(\frac{2\pi}{N}(k-j)\right)=
\left(\frac{N}{2}\right)^{N/4}.
\eeq
Substituting the square of this identity into (\ref{xc}) one finds
\beq
\psi_{\da\cdots\da}({\bf x})=(-1)^{\frac{N}{4}\left(\frac{N}{2}+1\right)}\left(\frac{N}{2}\right)^{N/2}. \label{daeq}
\eeq

Remember (\ref{esatta}) that
\beq
\psi_0({\bf x})=\left|\sum_F \psi_F({\bf x})\right|=N^{N/2} \label{somma}
\eeq
as we fixed the wave function to be real and positive.  As all $\psi_F$ have the same phase, this phase is inessential.  The key point is that $\psi_{\da\cdots\da}$ contributes $(N/2)^{N/2}$ of the total, which is $N^{N/2}$.  Stated differently, we may expand
\beq
{\rm{ln}}\left(\psi_0({\bf x})\right)=\alpha N^2 + \beta N\ {\rm{ln}}(N) + \gamma N + ... .
\eeq
The exact result is
\beq
\alpha=\gamma=0\hsp \beta=\frac{1}{2}.
\eeq
However just the even image $\psi_{\da\cdots\da}$ alone contributes
\beq
\alpha=0\hsp \beta=\frac{1}{2}\hsp \gamma=
-\frac{{\rm{ln}}(2)}{2}.
\eeq
Therefore $\psi_{\da\cdots\da}$ alone exactly yields the leading zero and nonzero terms, although the subleading term is incorrect.  This is no surprise, as the other terms in the sum (\ref{somma}), corresponding to other images, have not been included.

\subsection{Summing over Choices of Image: Arrows} \label{fsez}

In the rest of this section, we will include more images $F$ in an attempt to improve our estimate of $\gamma$.  Let us start by replacing the $i$th $\da$ with an $\ua$.    Now 
\beq
P_A^{\da\cdots\ua\cdots\da}(j)
=\left\{\begin{tabular}{ll}$2j$&{\rm{if\ }}$j\neq i$\\$2j-1$&{\rm{if}}\ $j=i$
\end{tabular}
\right.
\hsp P_B^{\da\cdots\ua\cdots\da}(j+N/2)=\left\{\begin{tabular}{ll}$2j-1$&{\rm{if\ }}$j\neq i$\\$2j$&{\rm{if}}\ $j=i$.
\end{tabular}\right.
\eeq

How does the expression (\ref{master}) for $\psi_{\da\cdots\ua\cdots\da}$ differ from $\psi_{\da\cdots\da}$?  In the latter case, the differences in the $P_A^F$ are all even.  Now, when $j=i$, $P_A(j)$ decreases by one unit and so the argument of the first sine decreases by $\pi/N$, while that of the second increases.  When $k=i$, the first increases and the second decreases.  As a result
\beq
\frac{\psi_{\da\cdots\ua\cdots\da}}{\psi_{\da\cdots\da}}=\prod_{j\neq i}^{N/2}\frac
{\s{\frac{2\pi}{N}\left(i-j-\frac{1}{2}\right)}\s{\frac{2\pi}{N}\left(i-j+\frac{1}{2}\right)}}{\st{\frac{2\pi}{N}\left(i-j\right)}}. \label{ijprod}
\eeq
So far our calculation has been exact.  Now let us consider an expansion around large $N$.  Choose a natural number $k<<N$.  Consider just those terms of the product such that $|i-j|\leq k$.  If the large $N$ limit is taken before the large $k$ limit, then this product converges in $k$ and we may approximate each sin$(x)$ by $x$ to obtain
\beq
\frac{\psi_{\da\cdots\ua\cdots\da}}{\psi_{\da\cdots\da}}\sim
\prod_{j=i-k}^{i+k}\frac{\left(i-j-\frac{1}{2}\right)\left(i-j+\frac{1}{2}\right)}{(i-j)^2}
=\left(\prod_{j=1}^{k}\frac{j^2-\frac{1}{4}}{j^2}\right)^2 \label{kprod}
\eeq
where the first product excludes $j=i$.  The first approximation, $k=1$, is
\beq
\frac{\psi_{\da\cdots\ua\cdots\da}}{\psi_{\da\cdots\da}}\sim\frac{9}{16}.
\eeq

The approximation $k=1$ is easy to understand.  It means that one only considers adjacent arrows.  Each pair corresponds to the following term
\beq
\da\da=\ua\ua=\st{\frac{2\pi}{N}}\hsp \da\ua=\ua\da=\s{\frac{\pi}{N}}\s{\frac{3\pi}{N}}. \label{kuno}
\eeq
Above we saw sequences of three arrows
\beq
\da\da\da={\rm{sin}}^4\left(\frac{2\pi}{N}\right)\hsp \da\ua\da=\st{\frac{\pi}{N}}\st{\frac{3\pi}{N}}
\eeq
and so we obtained the ratio
\beq
\frac{\psi_{\da\cdots\ua\cdots\da}}{\psi_{\da\cdots\da}}\sim\frac{\da\ua\da}{\da\da\da}\sim\frac{9}{16}.
\eeq

More generally, one may use the $k=1$ approximation (\ref{kuno}) to find $\psi_F$ for any $F$ that corresponds to a sequence of arrows, in other words any $F$ that includes precisely one member of each pair $\{2j-1,2j\}$.   The $k=1$ approximation is reasonable because $k=1$ yields the contribution to (\ref{kprod}) which is furthest from unity, and so the most important contribution to the absolute value of the matrix element.  This is the case because the corresponding $|i-j|=1$ term is the most important in (\ref{ijprod}).  But there is one exception to this observation.  If there is a $|i-j|=N/2-1$ term, it will have the same value.  Such a term only appears for the arrow at the beginning and at the end of $\mf$.  Therefore the $k=1$ approximation means that one considers only adjacent pairs of arrows in the $N/2$-vector $\mf$, where adjacency is understood modulo $N/2$ so that the first and last arrows are also adjacent.

 For an arbitrary string of arrows $\mf$, let $j$ be the number of occurances of $\da\ua$, with the above cyclicity condition understood.  Then $j$ will also be the number of occurances of $\ua\da$ and equivalently the number of strings of consecutive $\ua$'s and the number of strings of consecutive $\da$'s.  For example $\da\da\ua\ua\ua\ua\ua\da$ corresponds to $j=1$.  There are thus $2j$ subsequences $\ua\da$ or $\da\ua$, each leading to a factor of $3/4$ in the $k=1$ approximation according to Eq.~(\ref{kuno}).  Therefore
 \beq
 \frac{\psi_F}{\psi_{\da\cdots\da}}=\left(\frac{3}{4}\right)^{2j}.
 \eeq

How many $F$'s are there with each $j$?  One needs to place the $2j$ places where the direction of the arrow changes amongst $N/2$ arrows.  At each of these places one inserts $\ua\da$ or $\da\ua$, which is of length 2.  Thus there are a total of $N/2$ slots where one needs to place $2j$ transitions.  The number of such choices is ${N/2}\choose{2j}$.  Note that one can choose between $\ua\da$ and $\da\ua$ only once, leading to a single factor of $2$ which we will ignore.  For example, if the first transition is $\ua\da$ then the second will necessarily be $\da\ua$ and so on.  The corresponding contribution to $\psi_0$ is $\psi_1$ where
\bea
\frac{\psi_1}{\psi_{\da\cdots\da}}&=&
\sum_{j=0}^{N/4} {{N/2}\choose{2j}} \left(\frac{3}{4}\right)^{2j}
=\frac{1}{2}\left(
\sum_{j=0}^{N/2} {{N/2}\choose{j}} \left(\frac{3}{4}\right)^{j}
+\sum_{j=0}^{N/2} {{N/2}\choose{j}} \left(-\frac{3}{4}\right)^{j}\right)\nonumber\\
&=&
\frac{1}{2}\left(1+\frac{3}{4}\right)^{N/2}+\frac{1}{2}\left(1-\frac{3}{4}\right)^{N/2}\sim\left(\frac{7}{4}\right)^{N/2}. 
\eea
and so summing over all such $F$ and ignoring the overall phase we obtain
\beq
\psi_1=\left(\frac{7}{4}\right)^{N/2}\left(\frac{N}{2}\right)^{N/2}=\left(\frac{7N}{8}\right)^{N/2} \label{p1eq}
\eeq
or equivalently
\beq
\gamma=-\frac{{\rm{ln}}(8/7)}{2}.
\eeq

\subsection{Summing over Choices of Image: Beyond Arrows} \label{f4sez}

In general, not every pair $\{2j-1,2j\}$ will have precisely one element in $F$.  It may have $0$ or $2$ elements in $F$.  Let us include these cases by adding to our notation for pairs in (\ref{frdef})
\beq
2=\{1,1\}\hsp 0=\{0,0\}.
\eeq
With this short hand, any $N$-vector $\mf$, consisting of $0$'s and $1$'s, can be rewritten as an $(N/2)$-vector consistings of $0$'s, $\ua$'s, $\da$'s and $2$'s.   For example
\beq
\mf=\{0,1,1,1,1,0,0,0\}{\rm{\ corresponds\ to\ }}\da 2 \ua 0.
\eeq

What happens to the calculation in Subsec.~\ref{fsez} if we include a single adjacent pair $20$ or $02$ starting at the $i$th position?  Recall that in that subsection we adopted the approximation $k=1$ in which we only considered sine terms in (\ref{master}) corresponding to differences between neighboring symbols, understood modulo $N/2$ so that the first and last symbols are neighbors.   Let us refer to the sine terms that we consider as interactions.  In that case, there were $N/2$ symbols $\ua$ or $\da$, each of which had an interaction with the neighbor on each side leading to a total of $N/2$ interactions in $\ms_A$ and also in $\ms_B$, for a total of $N$ interactions.  Each interaction came with a factor of $1/N$, but these factors cancel in ratios $\psi_F$'s because there were the same number of interactions, $N$.

Now it is no longer obvious that the number of interactions is invariant when a $20$ replaces two arrows.  The two arrows were involved in 3 interactions: one with their neighbor on the left, one with their neighbor on the right and one between the two arrows.  When the two arrows are removed, so are these three interactions.  Now the $20$ has two interactions with the arrow on its left, and one interaction between the two arrows in the $2$.  So the number of interactions counted is the same when two arrows are replaced by a $20$.  This would not be the case were the $2$ and the $0$ separated.  Therefore, in the case of $20$ and $02$ insertions, we can continue to use our $k=1$ approximation in which only interactions between neighboring symbols are considered, but one must recall that there is now also an interaction inside of the $2$.

So what are the interactions of a $20$ with its neighbors?  The internal interaction gives the difference between the two elements of the $2$.  These are elements $2i-1$ and $2i$ so their difference is one, giving a factor of sin$\left(\frac{\pi}{N}\right)$ for both the $\ms_A$ and the $\ms_B$ terms in (\ref{master}).  The interactions with the neighbors depend on the value of the neighbor.  If the left neighbor is $\da$ then $2i-2\in F$ while $2i-3\in[1,N]\setminus F$.  The former is separated by $1$ unit from $2i-1$ and by $2$ units from $2i$, both of which are in $F$, and so one obtains a factor of sin$\left(\frac{\pi}{N}\right)$sin$\left(\frac{2\pi}{N}\right)$ from the $\ms_A$ terms.  If the left neighbor is $\ua$, one similarly obtains a factor of sin$\left(\frac{2\pi}{N}\right)$sin$\left(\frac{3\pi}{N}\right)$.  As neither $2i-1$ nor $2i$ is in $[1,N]\setminus F$, the left neighbor does not contribute to the $\ms_B$ terms.   The right hand neighbor similarly has two possibilities which lead to these two factors.  In all, we have three possible products of the six neighborly interactions involving the $20$
\bea
&&\s{\frac{\pi}{N}\left(P_A^F(i)-P_A^F(i-1)\right)}\s{\frac{\pi}{N}\left(P_A^F(i+1)-P_A^F(i-1)\right)}\nonumber\\
&&\times\ \s{\frac{\pi}{N}\left(P_A^F(i+1)-P_A^F(i)\right)}\s{\frac{\pi}{N}\left(P_B^F(i+1+N/2)-P_B^F(i+N/2)\right)}\nonumber\\
&&\times\ \s{\frac{\pi}{N}\left(P_B^F(i+2+N/2)-P_B^F(i+N/2)\right)}\nonumber\\
&&\times\s{\frac{\pi}{N}\left(P_B^F(i+2+N/2)-P_B^F(i+1+N/2)\right)}\nonumber\\
&&=\left\{\begin{tabular}{ll}
$\se{4}{\frac{\pi}{N}}\se{2}{\frac{2\pi}{N}}$&{\rm{\ for\ }}$\da 20\da$\\
$\se{3}{\frac{\pi}{N}}\se{2}{\frac{2\pi}{N}}\s{\frac{3\pi}{N}}$&{\rm{\ for\ }}$\da 20\ua$\ and $\ua 20\da$\\
$\se{2}{\frac{\pi}{N}}\se{2}{\frac{2\pi}{N}}\se{2}{\frac{3\pi}{N}}$&{\rm{\ for\ }}$\ua 20\ua$.\\
\end{tabular}
\right.
\eea
These can be compared with the $\se{6}{\frac{2\pi}{N}}$ that would be obtained for $\psi_{\ua\cdots \ua}$ or the average $(7/4)^2 \se{6}{\frac{2\pi}{N}}$ obtained for $\psi_1$ in Subsec.~\ref{fsez}.  

Again let us take the large $N$ limit by approximating sin$(x)\sim x$.  Then we obtain the ratios
\bea
&&\frac{\psi_{\da\cdots\da\da 20\da\da\cdots\da}}{\tilde{\psi}_1}=\frac{1^4 2^2}{\left(\frac{7}{4}\right)^2 2^6}=\frac{1}{49}\hsp
\frac{\psi_{\da\cdots\da\da 20\ua\da\cdots\da}}{\tilde{\psi}_1}=\frac{\psi_{\da\cdots\da\ua 20\da\da\cdots\da}}{\tilde{\psi}_1}=\frac{1^3 2^2 3^1}{\left(\frac{7}{4}\right)^2 2^6}=\frac{3}{49}\nonumber\\
&&\frac{\psi_{\da\cdots\da\ua 20\ua\da\cdots\da}}{\tilde{\psi}_1}
=\frac{9}{49}.
\eea
Here we have introduced the notation $\tilde{\psi}_1$ to denote the sum over 8 terms $\psi_{\da\cdots\da xxx\da\cdots\da}$ where $x=\da$ or $\ua$.  As a rough approximation, we will simply average these to obtain
\beq
\frac{\psi_{\cdots20\cdots}}{\psi_1}=\frac{4}{49}
\eeq
where $\psi_{\cdots20\cdots}$ is the sum over all $\psi$ with $\mf$ consisting entirely of arrows except for a single $20$ insertion at a fixed position.  The denominator is $\psi_1$ and not $\tilde{\psi}_1$ as we have now summed over all combinations of arrows even far from the $20$ insertion.

What about multiple insertions of $20$?  This is a length 4 string.  In principle it may be inserted at $N/2$ different places in the string of $N/2$ arrows, $0$'s and $2$'s.  However, many of these positions overlap.  To simplify the calculation, let us make the approximation that there are $N/4$ places to insert it, and simply add a degeneracy factor of $2$ reflecting the fact that it could be displaced by $1/2$ of a location, ignoring the possible overlaps.  Let us add another degeneracy factor of $2$ for the fact that $20$ may be replaced by $02$.  And let us ignore the fact that neighboring $20$'s and $02's$ will have different interactions than those treated above, where we assumed that the neighbors were arrows.  

With these approximations, the number of ways to insert $j$ strings $20$ is $N/4\choose j$ with a weight of $(4/49)^j$ and a degeneracy factor of $4^j$.  Altogether this yields the sum
\beq
\frac{\psi_2}{\psi_1}=\sum_{j=0}^{N/4}{N/4 \choose j}\left(\frac{16}{49}\right)^j=\left(\frac{65}{49}\right)^{N/4}=\left(\frac{\sqrt{65}}{7}\right)^{N/2} \label{p2eq}
\eeq
and so
\beq
\psi_2=\left(\frac{\sqrt{65}}{8}N\right)^{N/2}\hsp
\gamma=\frac{{\rm{ln}}(65)}{4}-\frac{{\rm{ln}}(8)}{2}.
\eeq
This is very close to the correct answer.  Of course there are further subdominant upwards corrections coming from adding separated pairs of $2$ and $0$, and further downwards corrections coming from higher $k$.  These corrections each shift $\gamma$ by a few percent.

\begin{table}
\centering
\begin{tabular}{c|c|c|c|}
&$\frac{1}{N}\left(\frac{\psi_{\da\cdots\da}}{\psi_0}\right)^{2/N}$&$\frac{1}{N}\left(\frac{\psi_1}{\psi_0}\right)^{2/N}$&$\frac{1}{N}\left(\frac{\psi_2}{\psi_0}\right)^{2/N}$\\
\hline\hline
Estimate Above&0.5&0.875&$1.008$\\
\hline
$N=6$&0.5&0.836&1\\
\hline
$N=8$&0.5&0.824&0.966\\
\hline
$N=10$&0.5&0.818&0.956\\
\hline
\end{tabular}
\caption{Comparison of large $N$, $k=1$ estimates above with an explicit Bethe Ansatz sum of the corresponding subsets of images $F$ at $N=6$, $N=8$ and $N=10$.  In the last three rows, the three columns sum over {\bf{(1)
}} $F=\da\cdots\da$, {\bf{(2)}} over all combinations of arrows and {\bf{(3)}} over all combinations of arrows with arbitrary insertions of $20$ and $02$.  Note that the first column agrees for all $N$ as $\psi_{\da\cdots\da}$ is calculated exactly.  When $N\leq 6$ all possible images $F$ either consist entirely of arrows or have a single $20$ or $02$ insertion, so the sum over all such images yields the full $S_N$.}
\label{untab}
\end{table}

In Table~\ref{untab} we compare the $k=1$ estimates above for the sums over subsets of the images $F$ with the exact sums over those subsets calculated using (\ref{bethe}) at $N=6$, $N=8$ and $N=10$.  The quantity $\psi_{\da\cdots\da}$ was calculated exactly above, for any $N$, and so it is of no surprise that (\ref{daeq}), reported in the first row of the first column, is equal to the exact results reported in the first column of the later rows.  $\psi_1$ on the other hand is systematically over estimated by (\ref{p1eq}), reported in the second column of the first row, with respect to the exact results in the second column of the later rows.  That is as a result of our $k=1$ approximation.  The $k=2$ correction, for example, multiplies our $(3/4)$ factor by $(15/16)$ yielding $45/64$ and so it reduces the expected $7/8$ to $109/64$.  Each $k$ correction in fact is negative.  However it is important to remember that they can only be applied to those insertions which are at least a distance $k$ from their neighbors.  This correction of course also applies to $\psi_2$, reported in the last column of the first row, as it was calculated from $\psi_2/\psi_1$ in (\ref{p2eq}).  With the $k=2$ correction, $\psi_2$ will be less than $\psi_0$, as it must be since it is a partial sum over positive contributions to $\psi_0$.  In our examples we also see that of course $\psi_2$ is less than $\psi_0$.   With more work, the estimates could become more precise.  However in the table we see that $\psi_2^{2/N}$ already provides an estimate of $\psi_0^{2/N}$ with an error of less than $5\%$.

\subsection{Summing over Choices of Image: A Systematic Approach}

The above system of ever increasing precision by summing over more possible images $F$ can be made systematic.  $P$ is a map from $[1,N]$ to $[1,N]$.  We always binned the domain into 2 bins, $\ms_A$ and $\ms_B$.   In the first step, in Subsec.~\ref{parisez}, we choose only the value of $F$ which gave the largest $\psi_F$, corresponding to all even or odd elements of $[1,N]$.  This gave us $\psi_{\ua\cdots\ua}=(N/2)^{N/2}$.  

Next in Subsec.~\ref{fsez} we also binned the range $[1,N]$, into $N/2$ bins $\{2j-1,2j\}$ and we only considered those images $F$ which contained precisely one value from each bin.  If $f_{Ij}(F)$ is the number of elements of $\ms_I$ that are mapped into $\{2j-1,2j\}$ then this corresponds to the case with all values equal to unity
\beq
f_{Ij}(F)=1. \label{bin1}
\eeq
We are trying to sum over all $g\in S_N$.  We have partitioned $S_N$ into cosets $H_F\subset S_N$ each labeled by $F$.  Now we have further partitioned $F$ into bins labeled by the $2\times N/2$ matrix $f_{Ij}$ with entries $0$, $1$ and $2$.   We considered only the sine terms, which we called interactions, in (\ref{master}) which connect neighboring bins $j$ and $j\p$.  Therefore in Subsec.~\ref{fsez} we have again considered the contributions to $\psi_0$ arising from the dominant bin, which now is (\ref{bin1}).  Whereas in Subsec.~\ref{parisez} the dominant bin, $\ua\cdots\ua$, contained $(N/2)!^2$ elements of the $N!$ elements in $S_N$, now the dominant bin (\ref{bin1}) includes permutations of the sets $\{2j-1,2j\}$ and so contains $2^{N/2}(N/2)!^2$ elements.  This bigger sum gave us a contribution of $\psi_1=(7N/8)^{N/2}$.

Actually one may make a similar interpretation of Subsec.~\ref{parisez} , where $\ms$ is the first row of a $2\times N$ matrix $f_{Ij}$ which gives the number of elements of $\ms_I$ equal to $j$, which is necessarily $0$ or $1$ as $j$ is a single element.  In this sense the difference between Subsecs.~\ref{fsez} and \ref{parisez} is that the first partitioned the image $[1,N]$ into $N/2$ bins while the second partitioned it into $N$ bins, so that each element was its own bin.

The next logical step would be to partition the image into $N/4$ bins and calculate the contribution of the dominant choice
\beq
f_{Ij}=2 \label{bin2}
\eeq
only considering interactions between neighboring bins.  We have not systematically done this.   However the condition (\ref{bin2}) corresponds to $20$, $02$, $\ua\ua$, $\ua\da$, $\da\ua$ and $\da\da$ which are the four combinations considered in Subsec.~\ref{f4sez}.  There our rough approximations, including strings $20$ and $02$, led us to the sum $\psi_1=(\sqrt{65} N/8)^{N/2}$, ever closer to the exact answer of $N^{N/2}$.  

\section{Two Densities} \label{doubsez}

We expect that the sum over permutations in the Bethe Ansatz (\ref{bethe}) is simpler when the $m_i$ or equivalently the $x_i$ are distributed linearly.  The purpose of binning is that while the $x_i$ are generally not linearly distributed for a matrix element of interest, by considering sufficiently small bins, the $x_i$ may be linear in each bin up to any desired precision.  

In Sec.~\ref{consez} the $x_i$ were already distributed linearly, as the density was taken to be constant.  The purpose of that section was merely to show how the binning approximation can be implemented systematically.  However it did not lead to any simplification, quite on the contrary.  

In this section we will therefore consider matrix elements of the Tonks-Girardeau ground state with states with two regions with distinct densities.   In Subsec.~\ref{vansez} one of these densities will be zero.  In Subsec.~\ref{stessosez}, each region will have the same number of particles.   In these cases, the Vandermonde determinant form for the ground state is of complexity $N^2$, and so in principle requires at least time $N^2$ to evaluate.  We will use the binning approach in a way which is manifestly independent of $N$, and so will arrive at results which may be a suitable starting place for a large $N$ limit.  

Of course in this case it would be possible to bin the Vandermonde product directly, and so bypass the complicated partitioning of the permutation group $S_N$.  However beyond the ground state in this model and also in other models treatable with the Bethe Ansatz, there is no Vandermonde product form, and so that method would have no hope of generalizing.

\subsection{When One Density Vanishes} \label{vansez}

In this case we will consider particle positions of the form
\beq
x_j=\frac{\alpha L}{N}j
\eeq
where $\alpha$ is an arbitrary real number.  Then (\ref{vanmat}) yields
\beq
\eta_j=e^{2\pi i x_j/L}=e^{2\pi i \alpha j/N}=\chi^{\alpha j/N}
\eeq
and the Vandermonde determinant of the matrix $\eta_j^k$ provides arbitrary matrix elements
\bea
\psi_0({\bf x})&=&\prod_{j<k}^{N}\left|\eta_k-\eta_j\right|=
2^{N(N-1)/2}\prod_{j<k}^N\left|\s{\frac{\pi \alpha}{N}\left(k-j\right)}\right|\nonumber\\
&=&2^{N(N-1)/2}\prod_{j=1}^{N-1}\left|\s{\frac{\pi \alpha}{N}j}\right|^{N-j}. \label{linvan}
\eea

This expression for the wave function is exact, but to calculate it requires a time polynomial in $N$ and so it is not suitable for a large $N$ limit.  To simplify it, we will need to choose a regime for $\alpha$.   The case $\alpha=1$ was the subject of Sec.~\ref{consez}.

\subsubsection{The Case: $0<\alpha<<1/2$} \label{apic}

In this case the argument of the sine function is always much less than $\pi/2$ and so we may use the approximation sin$(x)=x$ to obtain
\bea
\psi_0({\bf x})&\sim& 2^{N(N-1)/2}\prod_{j=1}^{N-1}\left(\frac{\pi \alpha}{N}j\right)^{N-j}
=\left(\frac{2\pi\alpha}{N}\right)^{N(N-1)/2}\prod_{j=1}^{N-1} \left(j!\right)\nonumber\\
&\sim&\left(\frac{2\pi\alpha}{N}\right)^{N(N-1)/2}\prod_{j=1}^{N-1}e^{j{\rm{ln}}\left(j\right)-j}
=\left(\frac{2\pi\alpha}{eN}\right)^{N(N-1)/2}{\rm{exp}}\left(\sum_{j=1}^{N-1}j{\rm{ln}}\left(j\right)\right)\nonumber\\
&\sim&\left(\frac{2\pi\alpha}{eN}\right)^{N(N-1)/2}{\rm{exp}}\left(\int_{j=1}^{N-1}dj\ j{\rm{ln}}\left(j\right)\right)\nonumber\\
&=&\left(\frac{2\pi\alpha}{eN}\right)^{N(N-1)/2}{\rm{exp}}\left(\frac{j^2}{2}{\rm{ln}}(j)-\frac{j^2}{4}\left|^N_1\right.\right)\nonumber\\
&\sim&\left(\frac{2\pi\alpha}{e^{3/2}}\right)^{N^2/2}. \label{pic}
\eea
Note that the leading term in the exponent, which is of order $N^2$ln$(N)$ in the penultimate line of (\ref{pic}), cancels in the final expression so long as $\alpha$ is held fixed in the large $N$ limit.

Even so, the norm of the state is of order $e^{N{\rm{ln}}(N)}$ and the volume of the $N$-dimensional coordinate space is of order $L^N$, so matrix elements of order $e^{-N^2}$ will have a measure zero contribution to the wave function even if they are all integrated over.  Thus states with $\alpha<<1/2$ do not contribute to the wave function in the large $N$ limit.

In Fig.~\ref{linafig} we compare our approximation (\ref{pic}) to the exact Vandermonde formula and find good agreement at large $N$ and $\alpha<0.5$.

\begin{figure} 
\begin{center}
\includegraphics[width=2.5in,height=1.7in]{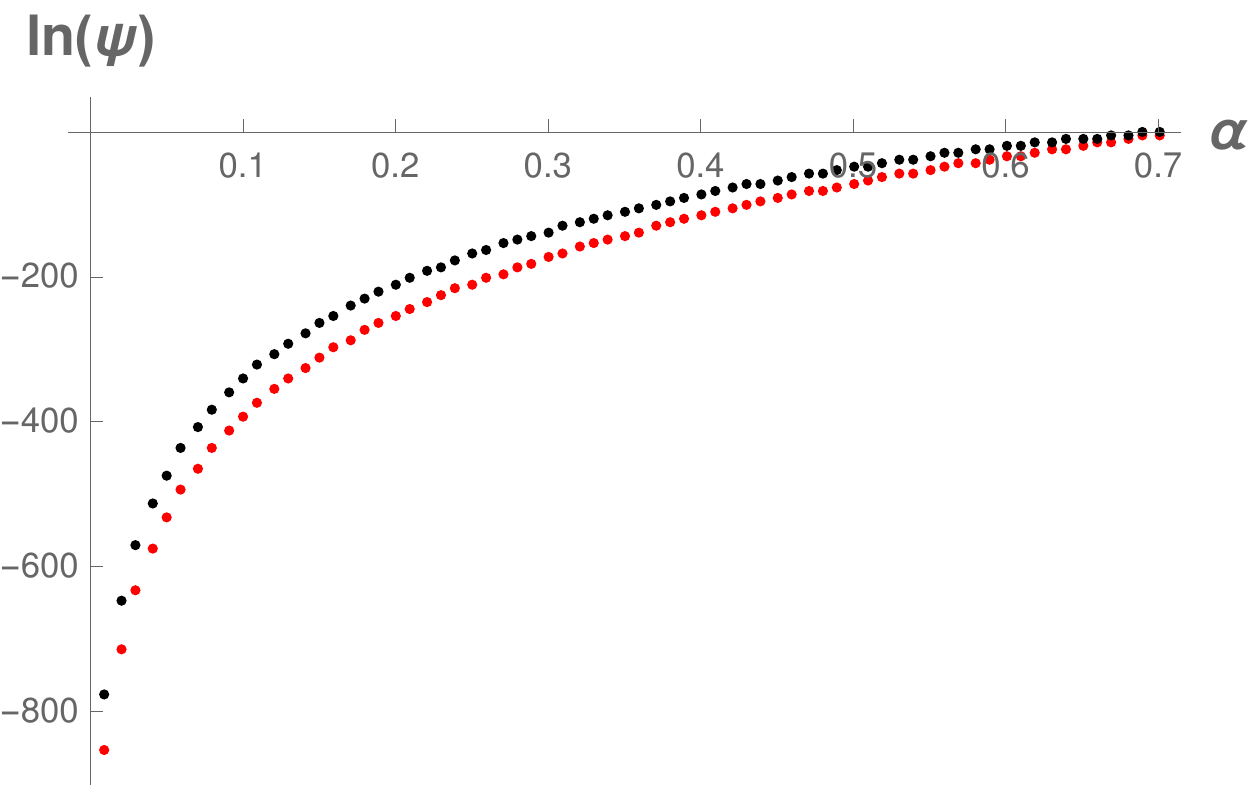}
\includegraphics[width=2.5in,height=1.7in]{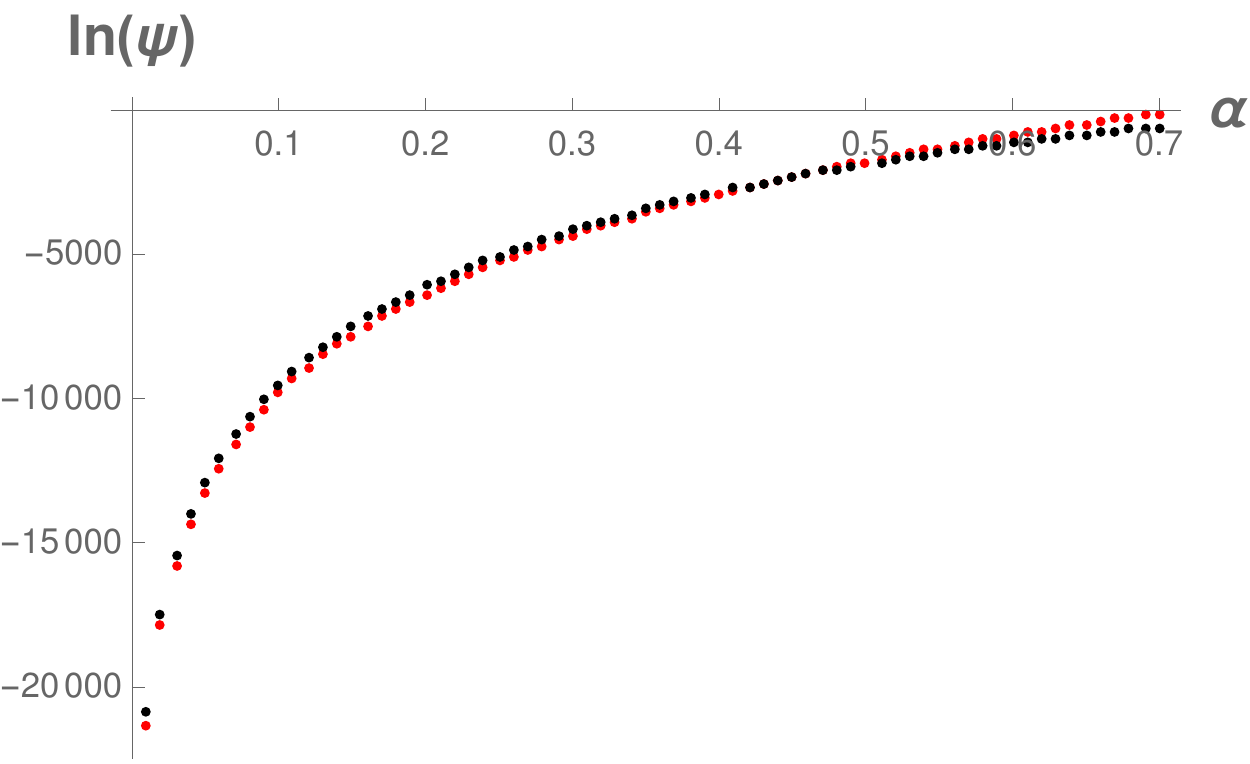}
\includegraphics[width=2.5in,height=1.7in]{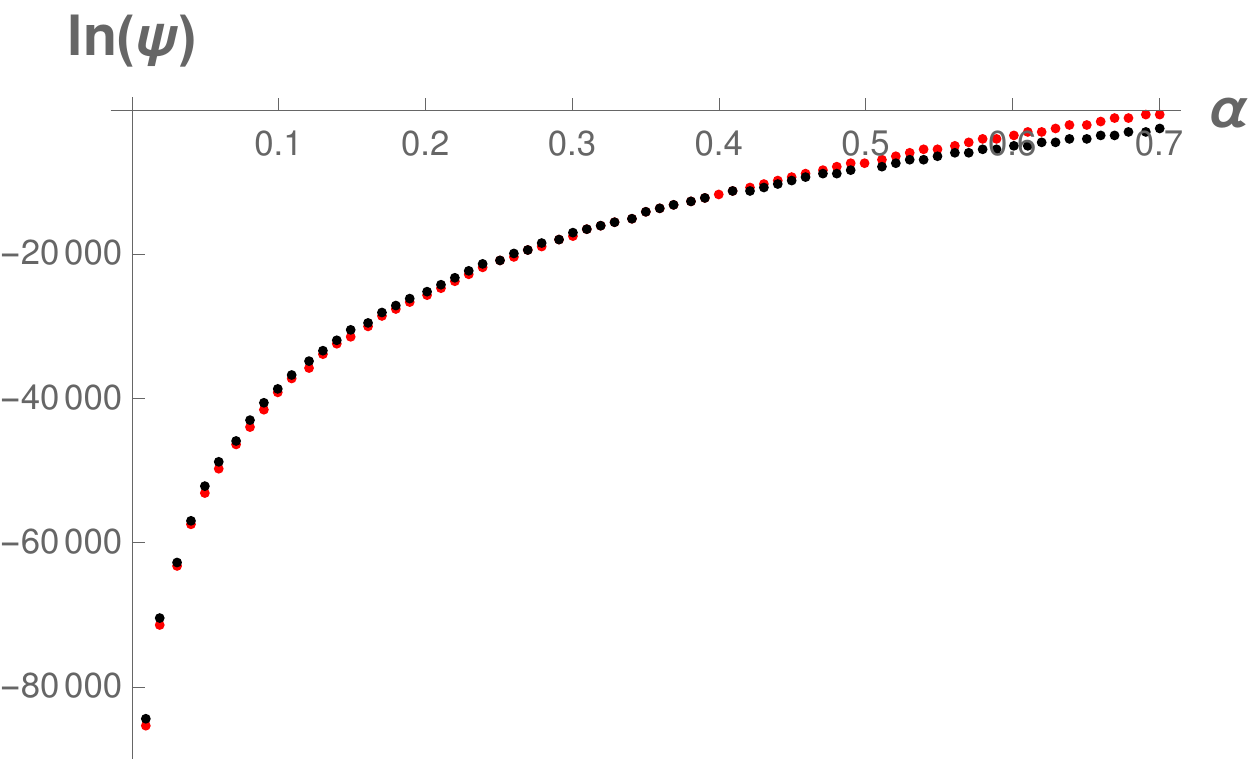}
\caption{Comparison of Eq.~(\ref{pic}) (red) versus the exact matrix elements calculated using the Vandermonde formula (\ref{linvan}) (black) at $N=20$ (top left), $N=100$ (top right) and $N=200$ (bottom). }
\label{linafig}
\end{center}
\end{figure}

\subsubsection{The Case: $1/2<<\alpha<1$}

We will make use of the identities
\beq
\prod_{j=1}^{M-1}\left(1-e^{\frac{2\pi i j}{M}}\right)=M \label{id1}
\eeq
and
\beq
\prod_{j=1}^{M-1}\left(1-e^{\frac{2\pi i j}{M}}\right)^{M-j}=M^{M/2} \label{id2}
\eeq
which were proved in Subsec.~\ref{essez}.  These allow us to use the fact that $\alpha$ is near $1$ to cleanly separate the $N^2$ terms in the exponent.

At large $N$, without loss of generality we can take $N/\alpha$ to be an integer.  Then
\beq
\psi_0({\bf x})=\left|\prod_{j=1}^{N-1}\left(1-\chi^{\alpha j}\right)^{N-j}\right|
=\left|\prod_{j=1}^{N/\alpha-1}\left(1-\chi^{\alpha j}\right)^{N-j}\right|/\left|\prod_{j=N}^{N/\alpha-1}\left(1-\chi^{\alpha j}\right)^{N-j}\right|=AB \label{ab}
\eeq
where $A$ and $B$ are defined to be the two factors in the penultimate term.  These can be evaluated separately
\bea
A&=&\left|\prod_{j=1}^{N/\alpha-1}\left(1-\chi^{\alpha j}\right)^{N-j}\right|=\left|\prod_{j=1}^{N/\alpha-1}\left(1-\chi^{\alpha j}\right)^{N/\alpha-j}\right|\left|\prod_{j=1}^{N/\alpha-1}\left(1-\chi^{\alpha j}\right)\right|^{N-N/\alpha}\nonumber\\
&=&\left(\frac{N}{\alpha}\right)^{N/(2\alpha)}\left(\frac{N}{\alpha}\right)^{N-N/\alpha}
=\left(\frac{N}{\alpha}\right)^{N\left(1-\frac{1}{2\alpha}\right)}
\eea
where the two factors in the first line were evaluated using the identities (\ref{id2}) and (\ref{id1}) respectively, with $M=N/\alpha$.  

Next
\bea
B&=&\left|\prod_{j=N}^{N/\alpha-1}\left(1-\chi^{\alpha j}\right)^{j-N}\right|=\left|\prod_{j=N-N/\alpha}^{-1}\left(1-\chi^{\alpha (j+N/\alpha)}\right)^{j-N+N/\alpha}\right|\\
&=&\left|\prod_{j=1}^{N/\alpha-N}\left(1-\chi^{\alpha j}\right)^{N/\alpha-N-j}\right|\sim\left|\prod_{j=1}^{N(1/\alpha-1)}\left(1-{\rm{exp}}\left(\frac{2\pi i (1-\alpha) j}{N(1/\alpha-1)}\right)\right)^{N(1/\alpha-1)-j}\right|.\nonumber
\eea
This is similar to $\psi_0$ in Subsec.~\ref{apic} but with
\beq
N\longrightarrow N\p=N(1-1/\alpha)\hsp\alpha\longrightarrow\alpha\p=1-\alpha
\eeq
so that $0<\alpha\p<<1/2$.    With these substitutions, Eq.~(\ref{pic}) yields
\beq
B\sim\left(\frac{2\pi(1-\alpha)}{e^{3/2}}\right)^{\frac{N^2\left(1-\frac{1}{\alpha}\right)^2}{2}}.
\eeq

Substituting $A$ and $B$ into Eq.~(\ref{ab}) we find the final result
\beq
\psi_0({\bf x})\sim \left(\frac{2\pi(1-\alpha)}{e^{3/2}}\right)^{\frac{N^2\left(1-\frac{1}{\alpha}\right)^2}{2}}\left(\frac{N}{\alpha}\right)^{N\left(1-\frac{1}{2\alpha}\right)}. \label{appb}
\eeq
As expected, when $\alpha=1$ at finite $N$, this expression equals (\ref{esatta}).  The limit of interest of course is $\alpha\longrightarrow 1$, $N\longrightarrow \infty$.  In this case, it differs from (\ref{esatta}) when the first term differs from unity, which is the case in which $N^2(1-\alpha)^2$ln$(1-\alpha)$ does not tend to $0$.

To see this, let us define
\beq
\epsilon=1-\alpha.
\eeq
The leading order terms in (\ref{appb}) as $\epsilon\longrightarrow 0$ are
\bea
{\rm{ln}}\left(\psi_0({\bf x})\right)
&\sim&\frac{N}{2}{\rm{ln}}(N)
+\frac{N^2\epsilon^2}{2}\left({\rm{ln}}(\epsilon)+{\rm{ln}}\left(\frac{2\pi}{e^{3/2}}\right)\right)
+\frac{N\epsilon}{2}\left({\rm{ln}}(N)+1\right)+\frac{N\epsilon^2}{2}\nonumber\\
&\sim&\frac{N}{2}{\rm{ln}}(N)+\frac{N^2\epsilon^2}{2}{\rm{ln}}(\epsilon)+\frac{N\epsilon}{2}{\rm{ln}}(N).
\eea
The leading term is just (\ref{esatta}).  The corrections will tend to $0$ in the large $N$ limit if $N\epsilon\rightarrow 0$, in other words if $\epsilon$ shrinks faster than $1/N$.  If $N\epsilon$ tends to a nonzero constant, then both constants are of the same order.  Otherwise, $N\epsilon\rightarrow\infty$, in which case the $N^2$ term dominates, corresponding to the first term in (\ref{appb}).     Which case is relevant?

Recall that the normalization of our energy eigenstates is of order $N!$ and so its logarithm is of order $N$ln$(N)$.  Also, if $x$ is discretized then the norms of the position states will have logarithms of order $N$.  Therefore we expect contributions from of order $e^N$ neighboring states, and terms in the logarithm of the matrix element of order $N$ and $N$ln$(N)$ may contribute to finite quantities.  For higher order terms however, no amount of normalization and integration can avoid the fact that one finds zero in the large $N$ limit.  This argument suggests that the dominant contributions to deformations will come from those that correct ${\rm{ln}}\left(\psi_0({\bf x})\right)$ with corrections of order $N$ or $N$ln$(N)$, and so correspond to
\beq
N^2\epsilon^2 {\rm{ln}}(\epsilon)\sim N
\eeq
and so
\beq
\epsilon\sim N^{-1/2}.
\eeq
In this case it is the first term in (\ref{appb}) which dominates.  This is fortunate, as we will soon see that the second is the most difficult to estimate in general.


In Fig.~\ref{linbfig} we compare our approximation (\ref{appb}) to the exact Vandermonde formula and again find good agreement at large $N$ and $0.5<\alpha<1$.

\begin{figure} 
\begin{center}
\includegraphics[width=2.5in,height=1.7in]{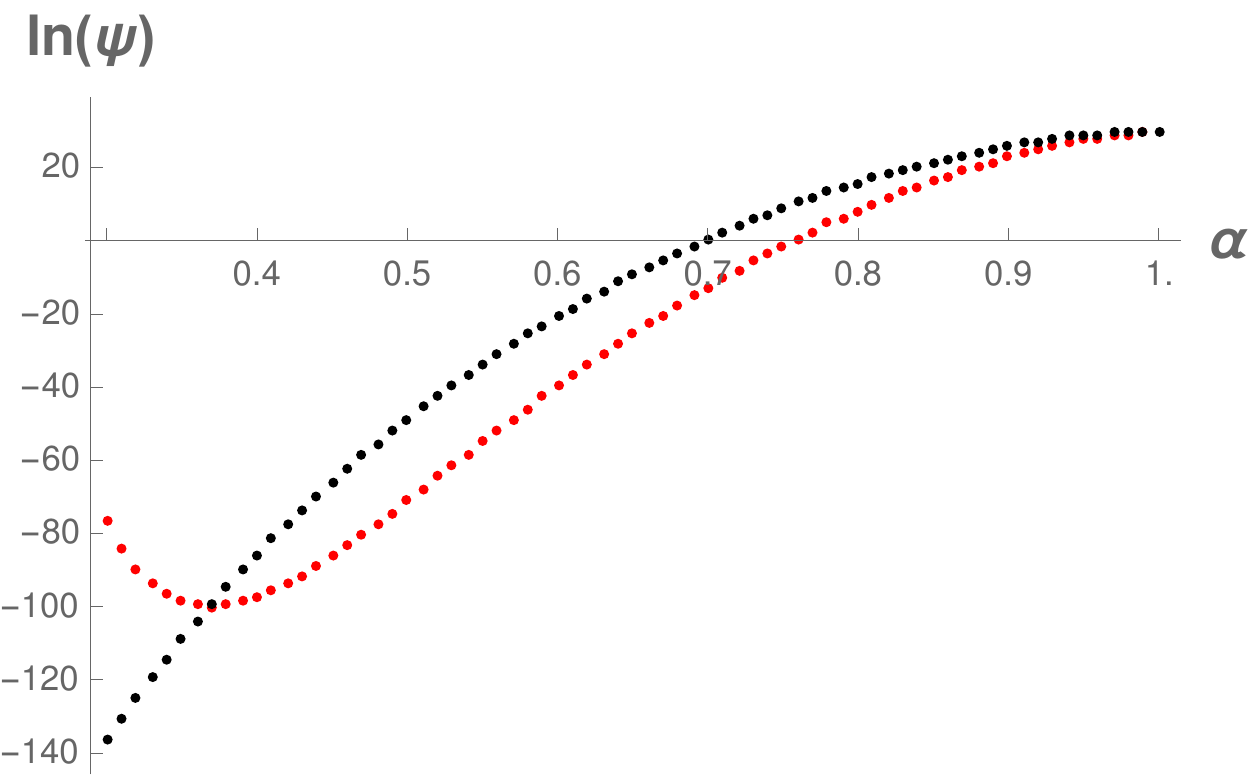}
\includegraphics[width=2.5in,height=1.7in]{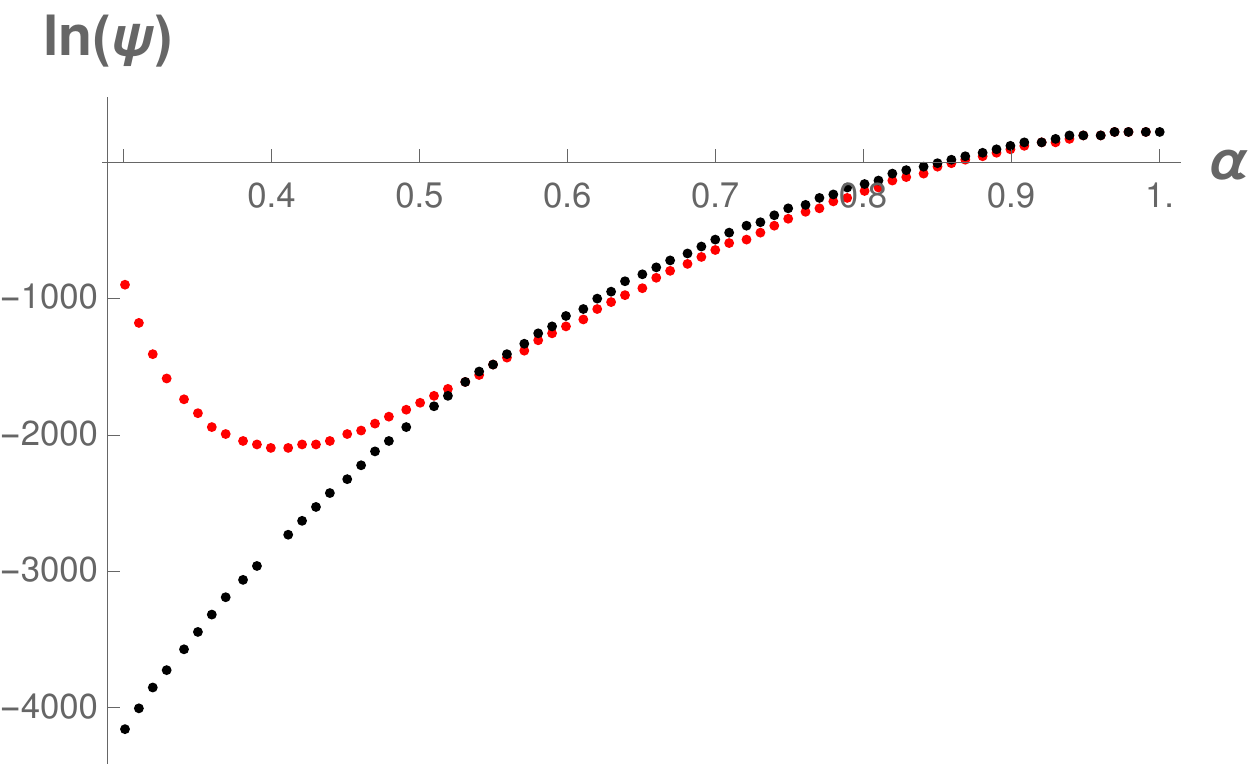}
\includegraphics[width=2.5in,height=1.7in]{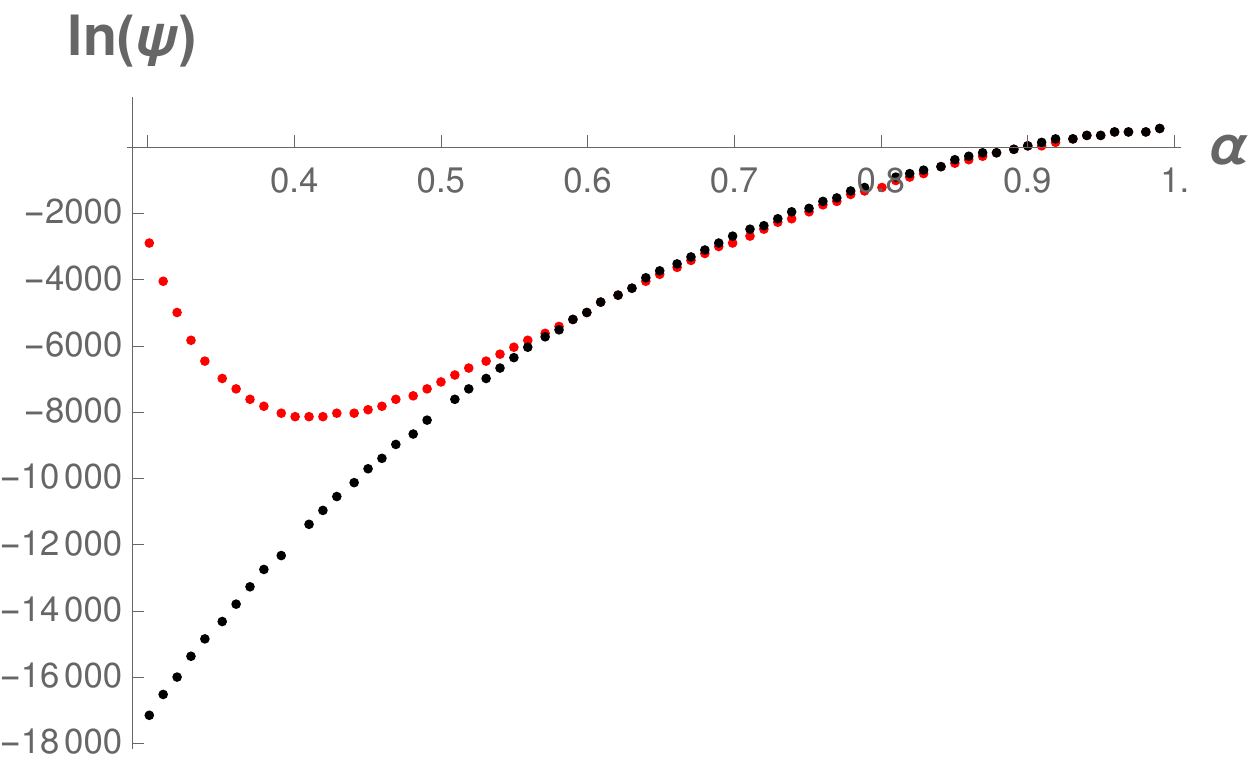}
\caption{Comparison of Eq.~(\ref{appb}) (red) versus the exact matrix elements calculated using the Vandermonde formula (\ref{linvan}) (black) at $N=20$ (top left), $N=100$ (top right) and $N=200$ (bottom). }
\label{linbfig}
\end{center}
\end{figure}

\subsubsection{The Case: $1<\alpha<<3/2$}

This is very similar to the previous case.  However now some signs change in the evaluation of $B$ and so we choose
\beq
N\longrightarrow N\p=N(1/\alpha-1)\hsp\alpha\longrightarrow\alpha\p=\alpha-1
\eeq
which lead to
\beq
\psi_0({\bf x})\sim \left(\frac{2\pi(\alpha-1)}{e^{3/2}}\right)^{\frac{N^2\left(1-\frac{1}{\alpha}\right)^2}{2}}\left(\frac{N}{\alpha}\right)^{N\left(1-\frac{1}{2\alpha}\right)}. \label{appc}
\eeq
Now the families of matrix elements which contribute to the wave function are neighborhoods of the cases in which $\alpha-1$ shrinks at least as quickly as $N^{-1/2}$.

In Fig.~\ref{lincfig} we compare our approximation (\ref{appc}) to the exact Vandermonde formula and again find reasonable agreement at large $N$ and $1<\alpha<1.3$.  The exact formula produces a series of sharp dips, corresponding to the values of $\alpha$ at which $\alpha j=N$ for some $j<N$.   Physically these are points where two particles coincide, which have zero wave function due to the impenetrability condition (\ref{imp}).  The Vandermonde formula has zeroes at
\beq
\alpha=\frac{N}{j}\hsp j<N. \label{zero}
\eeq
These zeroes are not captured by our approximation.  At large $N$ the zeroes become thin, but their density increases as $N$.  We do not know whether they persist in the large $N$ limit, or whether they lead to a physically relevant drop in the wave function at $\alpha>1$ in this limit.  However we will see below that in the two bin case, this regime does lead us to systematically overestimate matrix elements.

\begin{figure} 
\begin{center}
\includegraphics[width=2.5in,height=1.7in]{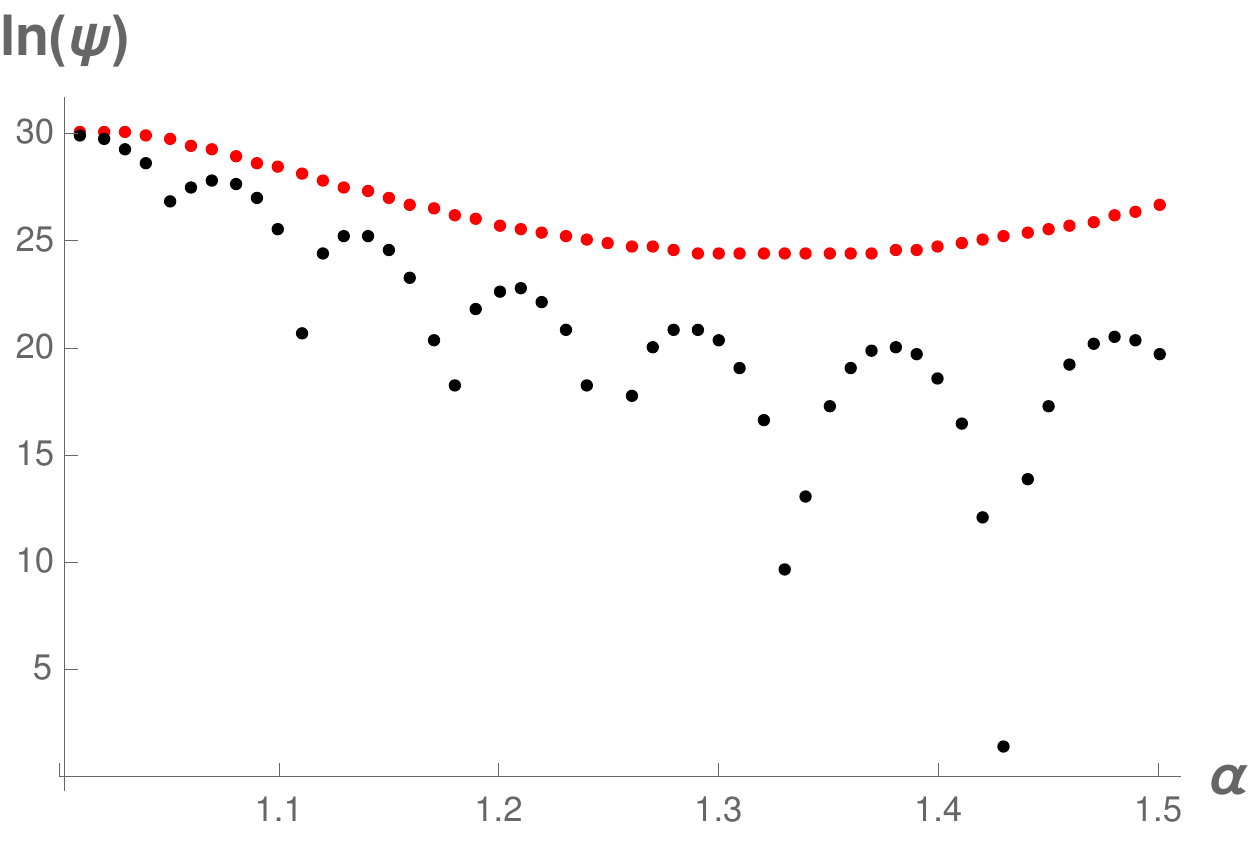}
\includegraphics[width=2.5in,height=1.7in]{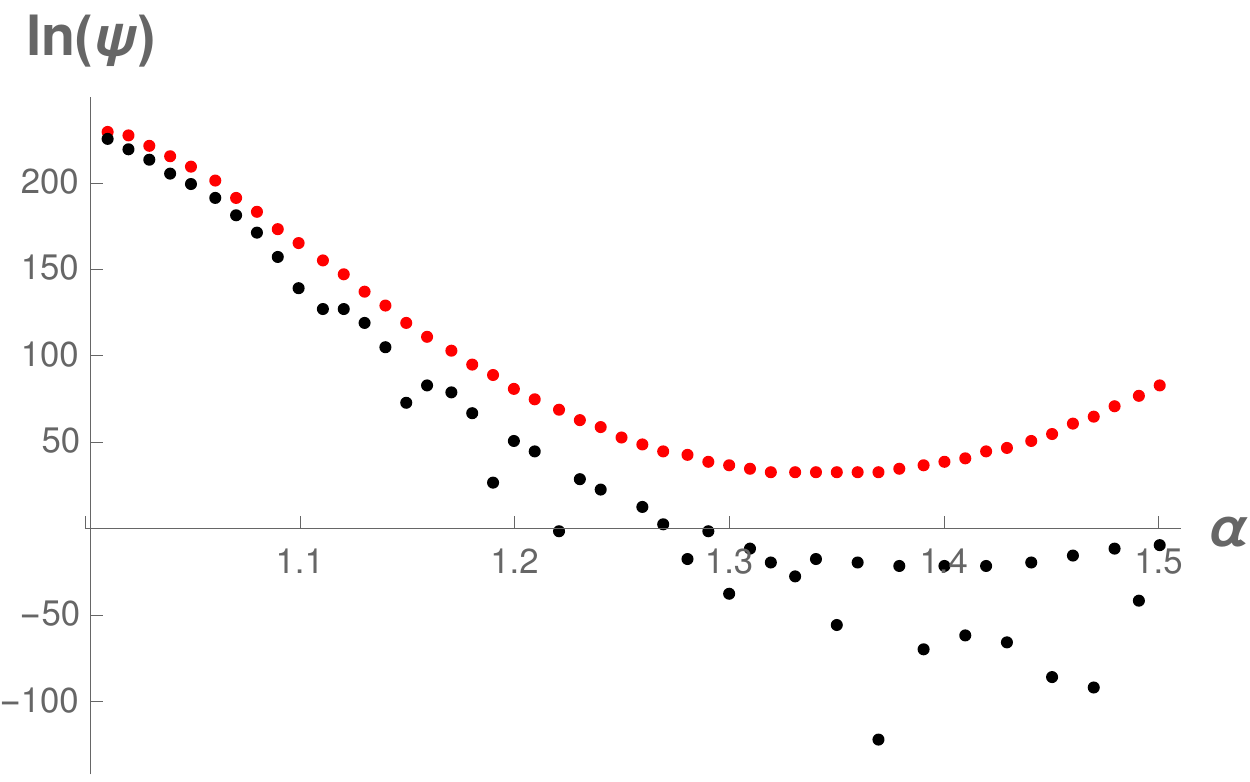}
\includegraphics[width=2.5in,height=1.7in]{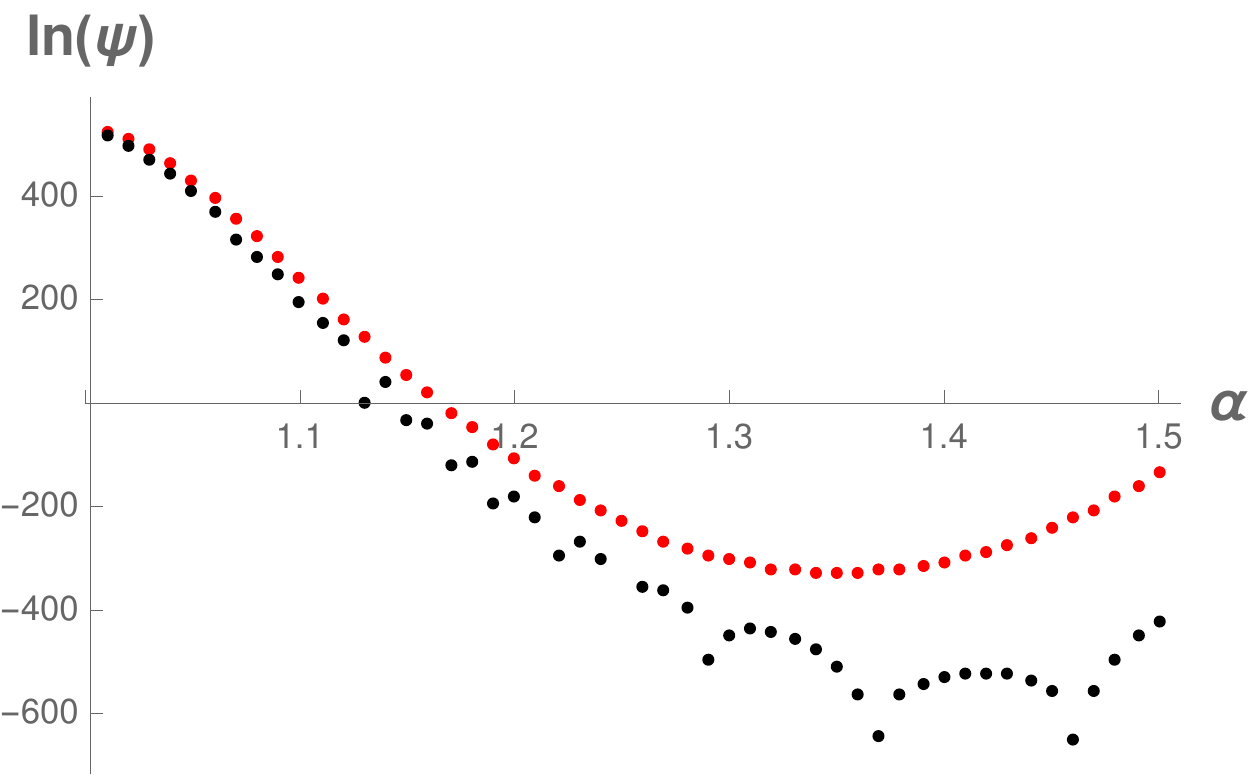}
\caption{Comparison of Eq.~(\ref{appc}) (red) versus the exact matrix elements calculated using the Vandermonde formula (\ref{linvan}) (black) at $N=20$ (top left), $N=100$ (top right) and $N=200$ (bottom). }
\label{lincfig}
\end{center}
\end{figure}

\subsection{Equal Number of Particles at Each Density} \label{stessosez}

Finally we are ready for a nontrivial application of the binning formalism.  We will consider an equal number of particles at density $\alpha$ and $2-\alpha$ in two contiguous regions
\beq
x_j=\left\{\begin{tabular}{ll}  $\frac{L}{N}\alpha j$&if\ $j\leq N/2$\\
$\frac{L}{N}(2-\alpha)j+L(\alpha-1)$&if\ $j\geq N/2$.
\end{tabular}\right. \label{2pos}
\eeq

Recall that our Bethe Ansatz (\ref{bethe}) describes the wave function in terms of a map $P:[1,N]\rightarrow [1,N]$ where the domain corresponds to our $x$'s and the range to the momenta, which are integers.  We will again consider the two bins $\ms_A$ and $\ms_B$ defined in Eq.~(\ref{msd}).  Our binning approach to the constant density case in Sec.~\ref{consez} suggests that we can systematically improve our estimates of the leading order term in ln$(\psi_0)$ by considering more and more images of $\ms_A$.   Our analysis of the constant density case in Subsec.~\ref{vansez} suggests that now this leading order will consist of the $N^2$ terms in ln$(\psi_0)$.

We now face new complications with respect to Sec.~\ref{consez}.  First, as the distance between some pairs of $x$ is greater than $L/N$, the argument of some of the sine functions in (\ref{master}) will be negative.  Second, there is no symmetry condition which determines which set $F$ should lead to the dominant contribution to the matrix elements.

As the $x_j$ in each bin have constant separation, the results of Subsec.~\ref{binmsez} largely carry over to this case.  The symmetric group $S_N$ can again be divided into cosets labeled by $F$,  defined such that $P(\ms_A)=F$ for all $g$ in the coset.  Then sum of (\ref{bethe}) over the coset yields a formula similar to Eq.~(\ref{passo4})
\bea
\psi_F({\bf x})&=&\sum_{g\in H_F}\exp{i\sum_{j=1}^N m(j)K_{P(j)}+\frac{i}{2}\sum_{j<k}^N\Psi_{P(j),P(k)}}\nonumber\\
&=& \sum_{g\in H_F} (-1)^g \chi^{\alpha\sum_{j=1}^{N/2} j P(j)+\sum_{j=N/2+1}^{N}\left((2-\alpha)j+(\alpha-1){N}\right)P(j)} .
\eea
The average value of $x$ in $\ms_A$ still differs from that of $\ms_B$ by $L/2$.  Therefore again, up to an overall phase, $\psi_F$ can be written as a product $\psi_A\psi_B$.  

As the $x_j$ are linear in each bin, these can again be written as Vandermonde determinant and so as a product of sines, now with factors of $\alpha$ and $2-\alpha$ respectively.  Recombining them we obtain the analogue of Eq.~(\ref{master})
\beq
\left|\psi_F({\bf x})\right|= 2^{\frac{N}{2}\left(\frac{N}{2}-1\right)} \prod_{j<k}^{N/2}{\rm{sin}}\left(\frac{\pi\alpha}{N}\left(P_A^F(k)-P_A^F(j)\right)\right){\rm{sin}}\left(\frac{\pi(2-\alpha)}{N}\left(P_B^F(k+N/2)-P_B^F(j+N/2)\right)\right) . \label{master2}
\eeq
Again the phases are $F$-independent except for the sines themselves, which are now negative if their arguments are greater than $\pi$.  Therefore the $\psi_F$ will all contribute to $\psi_0$ with the same phase up to a sign.  In principle we could calculate the distribution of these signs to refine our estimate of $\psi_0$, however we will leave this to future work.

The zeros and sign changes of the sine arise identically to those in Eq.~(\ref{zero}) in the case of a constant density.  In that case they had a very physical origin, they were the zeroes of the wave function $\psi_0$ when two particles overlap.  Now the particle positions are given by (\ref{2pos}) and no particles overlap, therefore the total wave function $\psi_0$ does not vanish.

Instead the zeroes in (\ref{master2}) are zeroes in $\psi_F$, corresponding to the sum over a single image $F$.  These zeroes come from the fact that the wave function would have vanished had the same $x$ spacing $\alpha$ or $2-\alpha$ persisted over all of $[1,N]$ instead of just the bin $\ms_A$ or $\ms_B$.   More concretely, they arise from the fact that the wave function Slater determinant contains all behaviors from $e^{-i(N-1)x/2L}$ to $e^{i(N-1)x/2L}$ and so if two adjacent $x$'s differ by $L/(N-1)$, for example, corresponding to $\alpha=N/(N-1)$, then a minor in the determinant which includes $e^{-i(N-1)x/2L}$ for one $x_j$ and $e^{i(N-1)x/2L}$ for its neighbor will vanish.   In this sense, the zeroes of (\ref{master2}) are zeroes in a minor and not in the determinant so they are spurious and it would be desirable if (\ref{bethe}) could be resummed in such a way that they are not present from the beginning.

Our goal will now be to make the crudest possible estimate of $\psi_0$, as a $\psi_F$ which is in a sense dominant.  What do we mean by dominant?  Different definitions give different estimates.  One choice of image is just $F=\da\cdots\da$ as before.  In Fig.~\ref{2fig} this choice is compared with the exact $\psi_0$ calculated using (\ref{vanmat}).

\begin{figure} 
\begin{center}
\includegraphics[width=2.5in,height=1.7in]{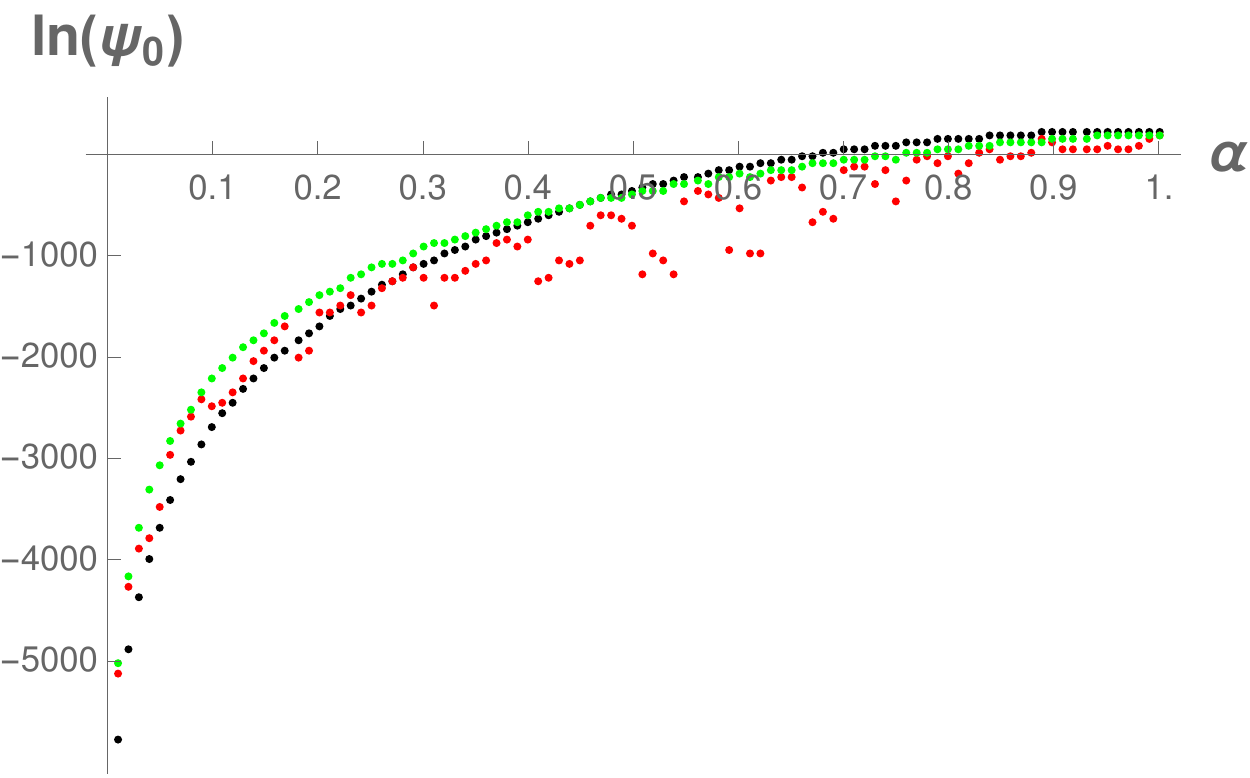}
\includegraphics[width=2.5in,height=1.7in]{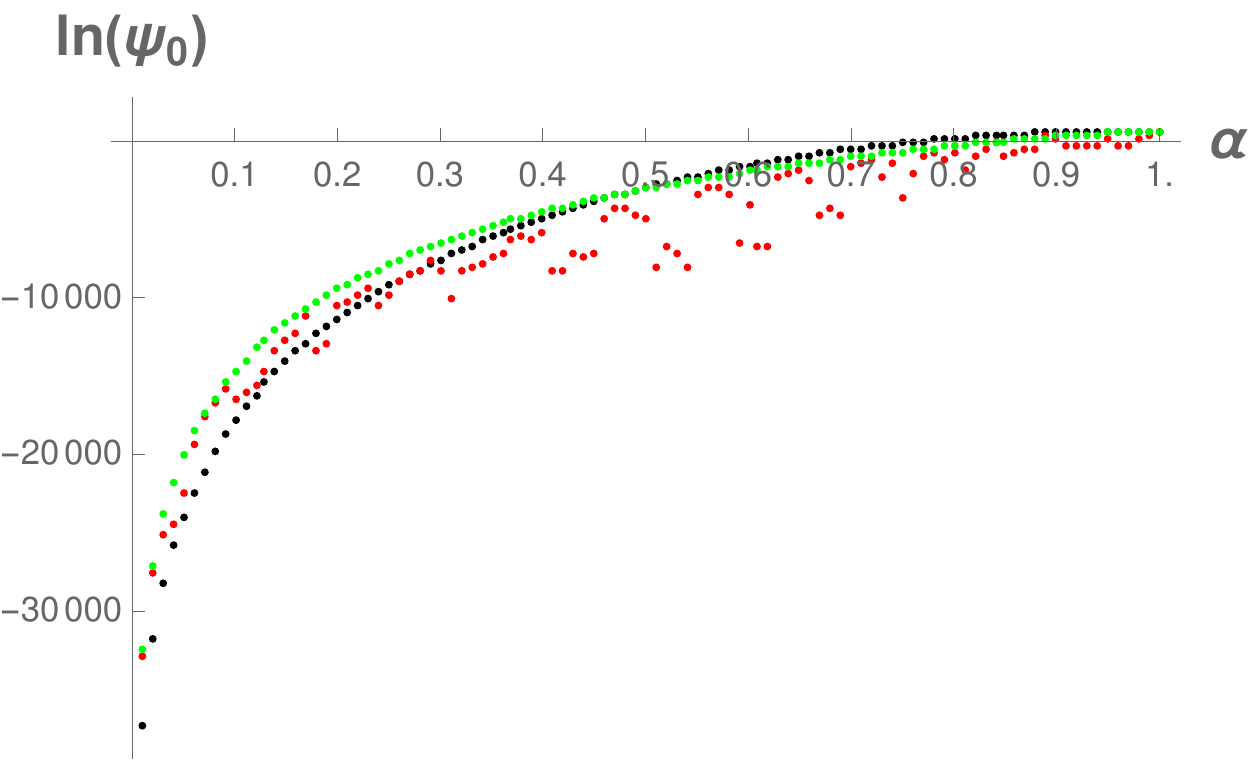}
\includegraphics[width=2.5in,height=1.7in]{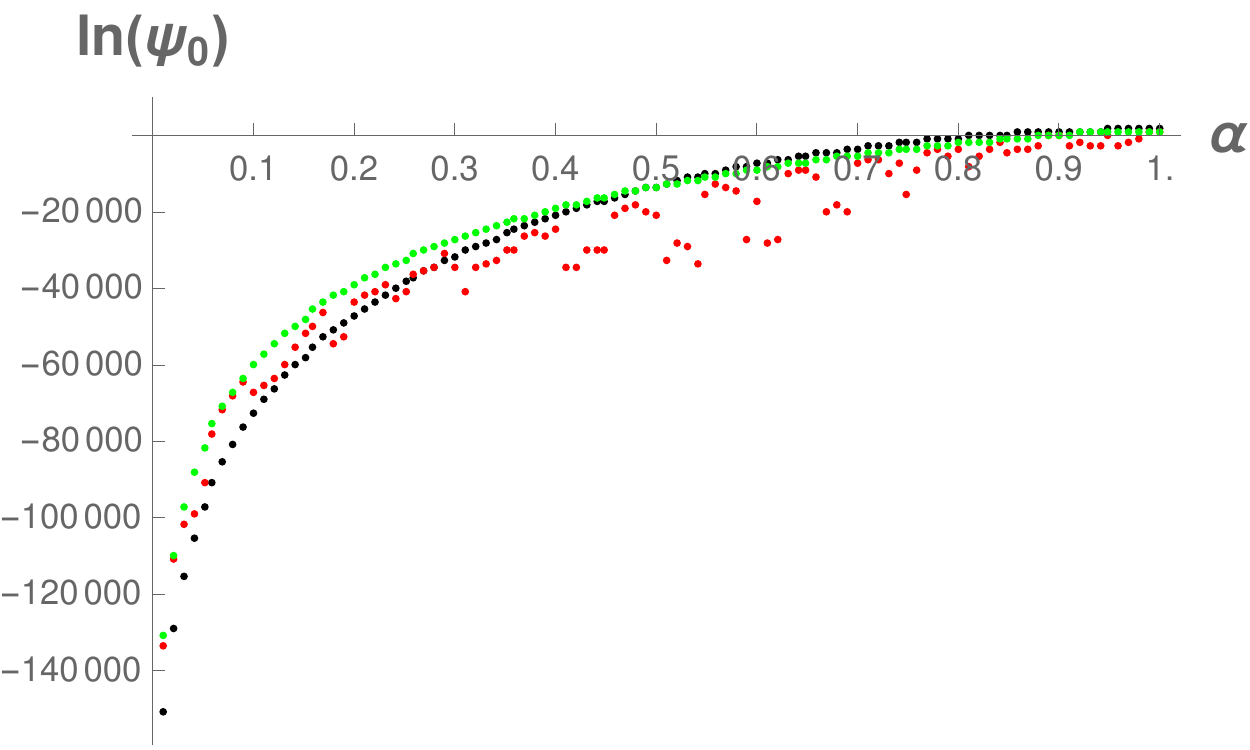}
\includegraphics[width=2.5in,height=1.7in]{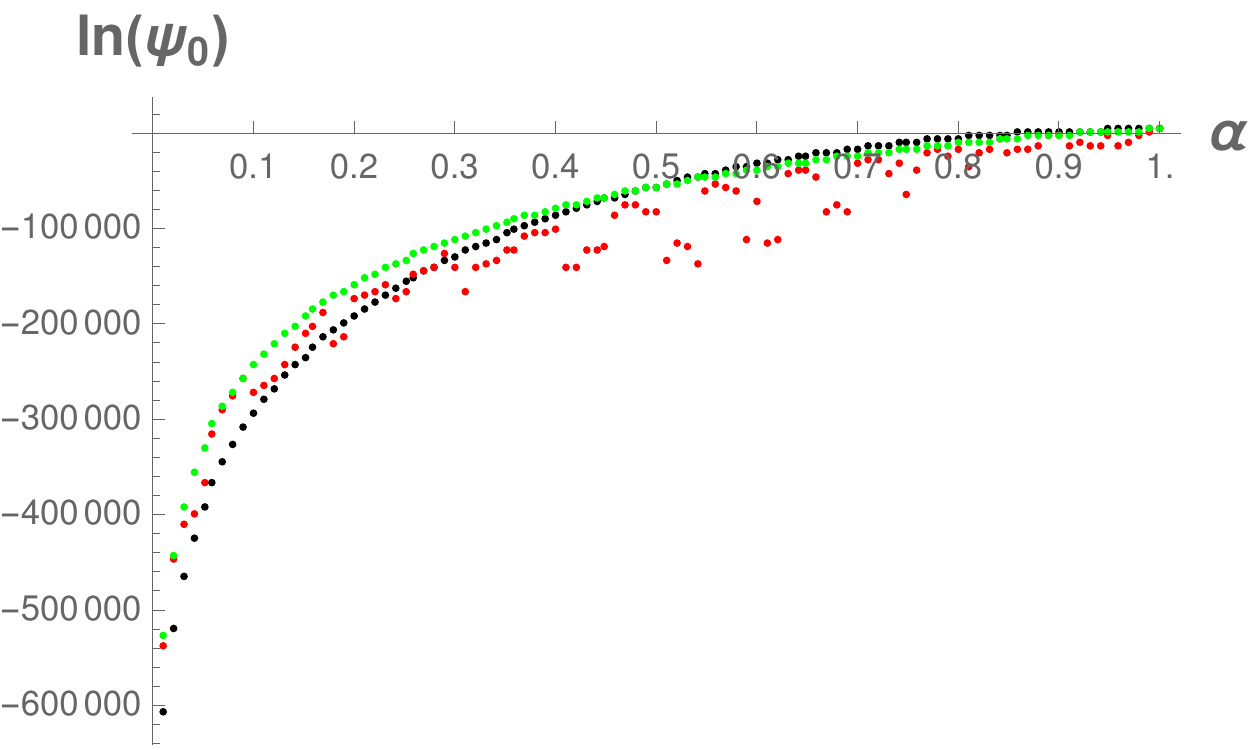}
\includegraphics[width=2.5in,height=1.7in]{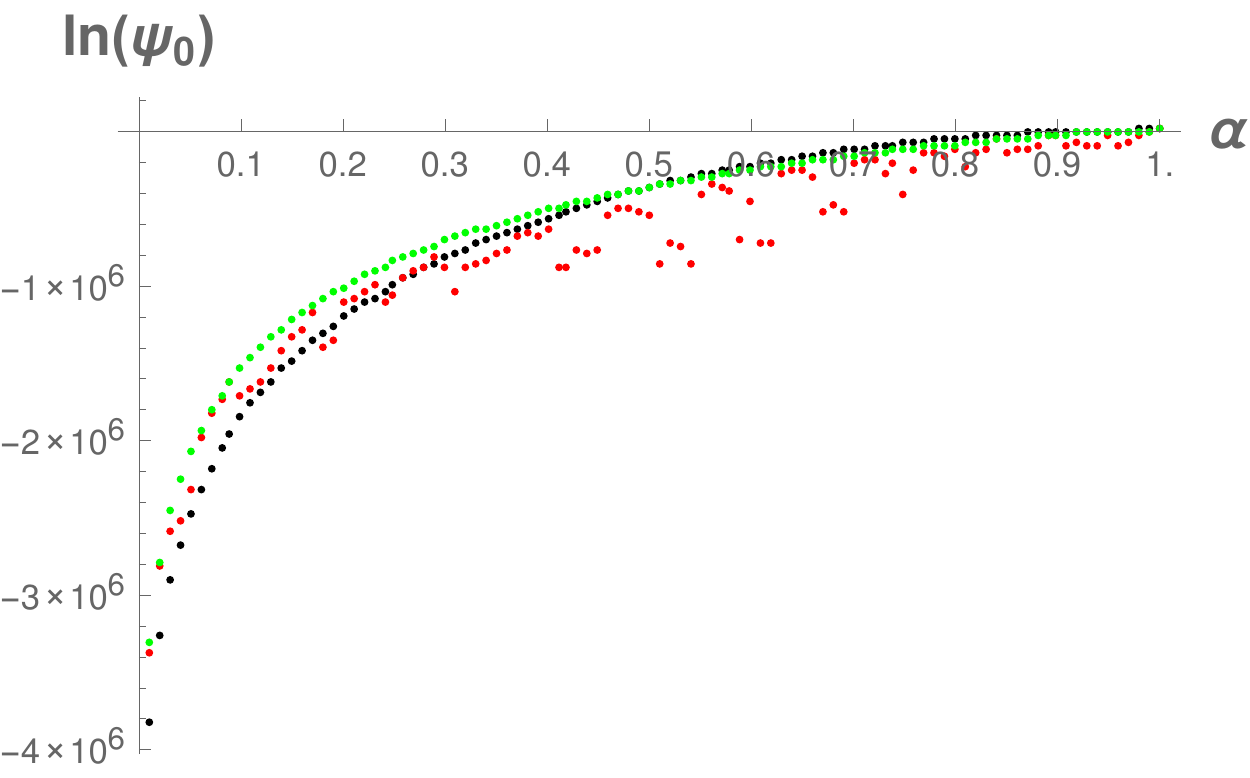}
\includegraphics[width=2.5in,height=1.7in]{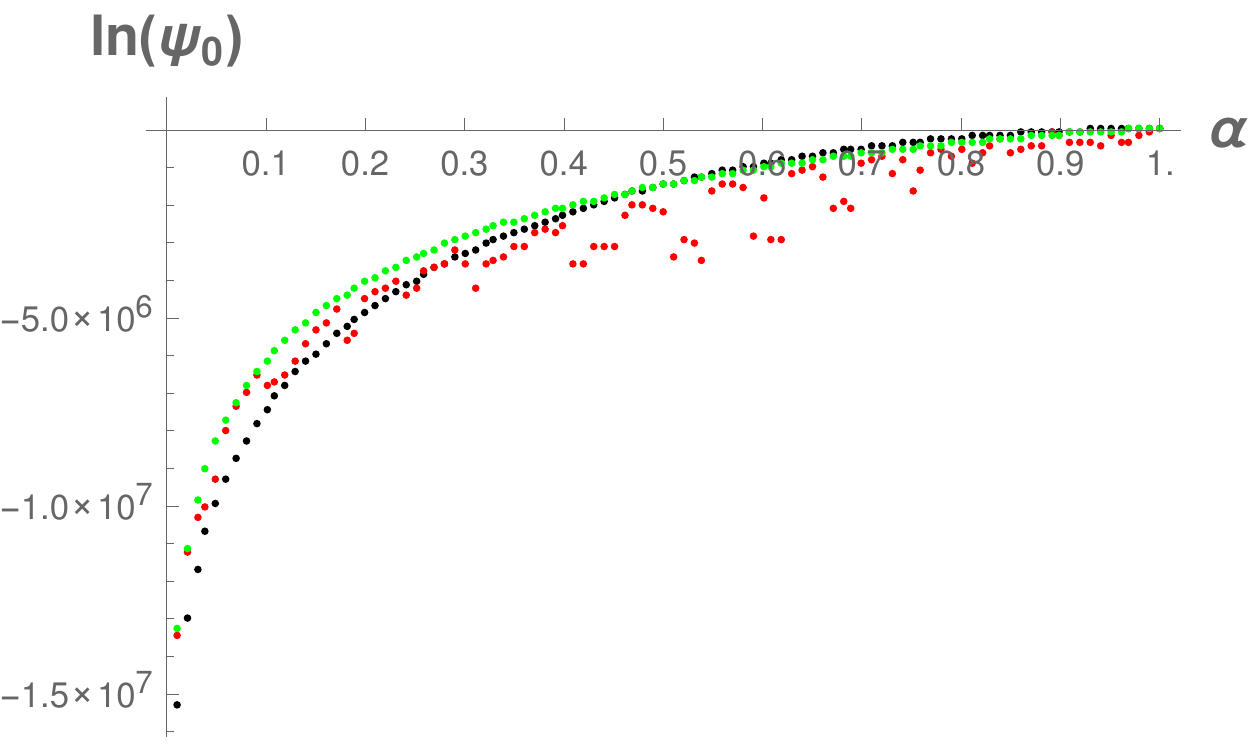}
\caption{Comparison of the estimator in Subsec.~\ref{optsez} (red) versus ln$(\psi_{\da\cdots\da})$ (green) and the exact matrix elements ln$(\psi_0)$ calculated using the Vandermonde formula (\ref{vanmat}) (black) at $N=100$ (top left), $N=250$ (top right), $N=500$ (middle left), $N=1000$ (middle right), $N=2500$ (bottom left) and $N=5000$ (bottom right). }
\label{2fig}
\end{center}
\end{figure}

\subsection{Optimizing the Image} \label{optsez}

How does our estimate of $\psi_0$ depend upon our choice of $F$.  Recall that $F$ is a subset of $[1,N]$ consisting of $N/2$ elements.  We can parameterize this interval by
\beq
x=\frac{i}{N}\hsp i\in [1,N]
\eeq
and in the large $N$ limit we can associate a local density
\beq
0\leq\rho(x)\leq 1
\eeq
with the subset $F\in[1,N]$.
 
Clearly $F$ determines $\psi_F$.  And $F$ determines $\rho(F)$.  But to what extent does $\rho(F)$ determine $\psi_F$?  We claim that $\rho(F)$ fixes the dominant $N^2$ term in ln$(\psi_0)$ and also perhaps the $N$ln$(N)$ term.  We cannot prove this claim, but we will provide an example.

Consider two choices of image $F$.  One is $F_1=F_{\da\cdots\da}$ which is the subset of even elements.  The second, $F_2=F_{\da\ua\da\cdots\ua}$ alternates between $\ua$ and $\da$, corresponding to $\mf=\{011001100\cdots01\}$.   For simplicity we will set $\alpha=1$, so that $\alpha$ does not dominate over the effect of this distinction.  We do not expect that this choice will affect our conclusions.  In both cases the separations between elements of $F$ and $[1,N]\setminus F$ are equal, and so $\psi_A=\psi_B$.

In the first case
\bea
\left|\psi_{F_1}({\bf x})\right|&=& 2^{\frac{N}{2}\left(\frac{N}{2}-1\right)} \prod_{j<k}^{N/2}{\rm{sin}}\left(\frac{\pi}{N}\left(P_A^F(k)-P_A^F(j)\right)\right)^2 \nonumber\\
&=&2^{\frac{N}{2}\left(\frac{N}{2}-1\right)} \prod_{j<k}^{N/2}{\rm{sin}}\left(\frac{2\pi}{N}\left(k-j\right)\right)^2=\left(\frac{N}{2}\right)^{N/2}. \label{casea}
\eea
and in the second
\bea
\left|\psi_{F_2}({\bf x})\right|&=&2^{\frac{N}{2}\left(\frac{N}{2}-1\right)} \prod_{k=1}^{N/4-1}{\rm{sin}}^{N/4}\left(\frac{\pi}{N}\left(4k-3\right)\right){\rm{sin}}^{N/4}\left(\frac{\pi}{N}\left(4k-1\right)\right){\rm{sin}}^{N/2}\left(\frac{\pi}{N}\left(4k\right)\right)\nonumber\\&=&2^{\frac{N}{2}\left(\frac{N}{2}-1\right)} 
\left(\prod_{k=1}^{N-1}{\rm{sin}}\left(\frac{\pi}{N}k\right)\right)^{N/4}
\left(\prod_{k=1}^{N/2-1}{\rm{sin}}\left(\frac{2\pi}{N}k\right)\right)^{-N/4}
\left(\prod_{k=1}^{N/4-1}{\rm{sin}}\left(\frac{4\pi}{N}k\right)\right)^{N/2}\nonumber\\
&=&N^{N/4}\left(\frac{N}{2}\right)^{-N/4}\left(\frac{N}{4}\right)^{N/2}=\left(\frac{N}{2\sqrt{2}}\right)^{N/2}. \label{caseb}
\eea
In this case the change in local density only affects the $O(N)$ term in ln$(\psi)$.

Thus the leading behavior of ln$(\psi)$ appears to be entirely determined by $\rho(x)$.  How do we optimize $\rho(x)$?  Let us take $\psi_A$ and $\psi_B$ to be real and positive.  We will now allow $\alpha$ to be an arbitrary real number.  In this case the arguments of the sine terms in (\ref{casea}) and (\ref{caseb}) do not necessarily lie in the interval $[0,\pi]$.  However the absolute value on the left hand side of these equations means that each sine term should be $|\rm{sin}|$.  Then
\bea
\psi_A&=&2^{\frac{N}{4}\left(\frac{N}{2}-1\right)}{\rm{Exp}}\left(\sum_{j<k}^{N/2}{\rm{ln}}\left|\s{\frac{\pi\alpha}{N}\left(P_A^F(k)-P_A^F(j)\right|}\right)\right)=2^{\frac{N}{4}\left(\frac{N}{2}-1\right)} e^{c_A} \label{disceq}\\
\psi_B&=&2^{\frac{N}{4}\left(\frac{N}{2}-1\right)}{\rm{Exp}}\left(\sum_{j<k}^{N/2}{\rm{ln}}\left|\s{\frac{\pi(2-\alpha)}{N}\left(P_B^F(k+N/2)-P_B^F(j+N/2)\right|}\right)\right)\nonumber\\&=&2^{\frac{N}{4}\left(\frac{N}{2}-1\right)} e^{c_B}\nonumber
\eea
where $c_A$ and $c_B$ are defined by the preceding expressions.  In the continuum limit the sums are replaced by integrals over $x$ and $y$ and the set $F$ is replaced by its density  $\rho(x)$
\bea
c_A&\sim& \int_0^1 dx \int_{x+\epsilon}^1 dy \rho(x)\rho(y)  {\rm{ln}}\left|\s{\pi\alpha\left(y-x\right)}\right|\\
c_B&\sim& \int_0^1 dx \int_{x+\epsilon}^1 dy (1-\rho(x))(1-\rho(y))  {\rm{ln}}\left|\s{\pi(2-\alpha)\left(y-x\right)}\right|\nonumber
\eea
where $\epsilon$ is a small number that cuts off the divergence as $x\rightarrow y$, reflecting the fact that $j\neq k$ in the original discrete expression (\ref{disceq}).  The expressions $c_A$ and $c_B$ are those for the energy of a system of density $\rho$ with an interparticle potential given by ln(sin).

\begin{figure} 
\begin{center}
\includegraphics[width=2.5in,height=1.7in]{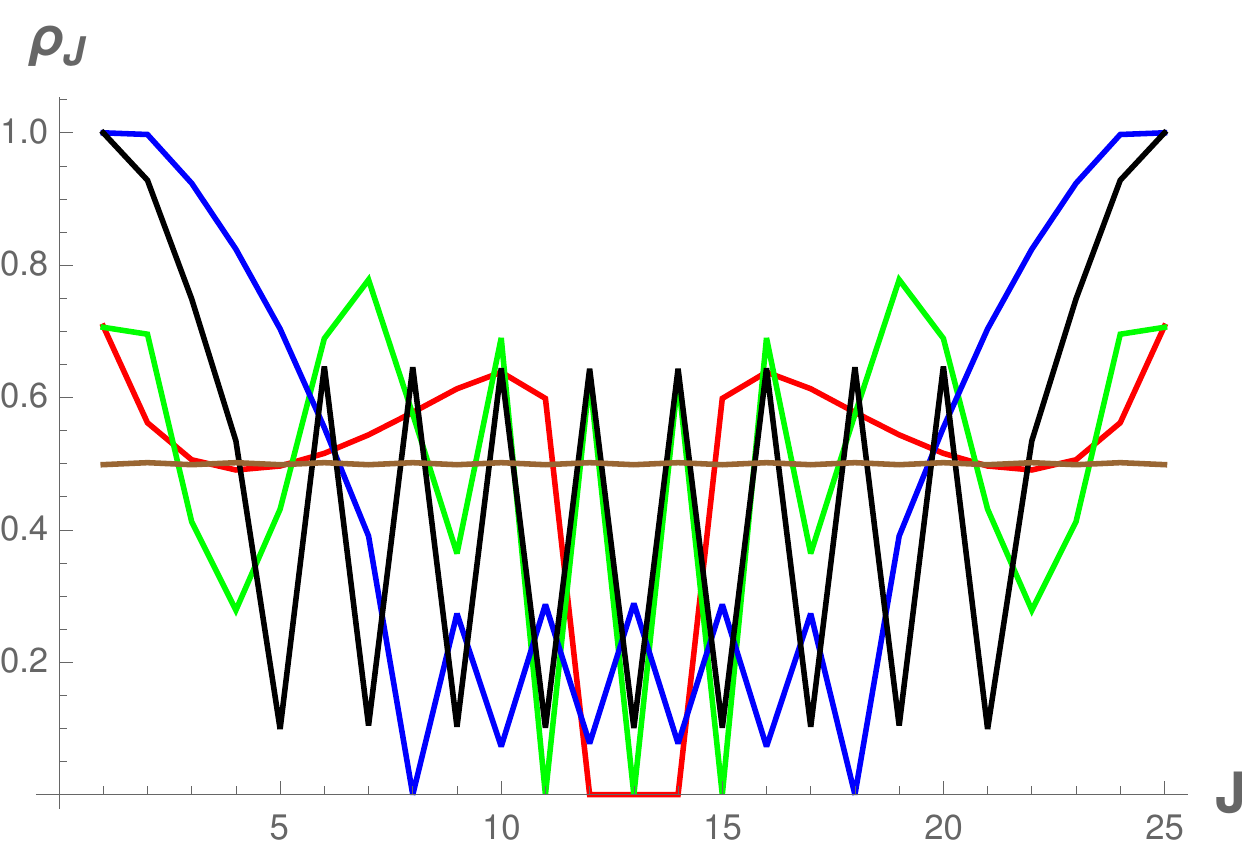}
\caption{The optimal values of $\rho$ found by solving (\ref{appeq}) discretized into 25 parts $J$.   The red, green, blue, black and brown curves correspond to $\alpha$ equal to $0.2,\ 0.4,\ 0.6,\ 0.8$ and $1$ respectively.  Note that when $\alpha=1$ the optimal density is constant as expected.}
\label{rhofig}
\end{center}
\end{figure}

In this analogy, the potential $V(z)$ for a particle at $x$ is the functional derivative with respect to $\rho(z)$ of $c_A+c_B$.  However $F$ has a fixed number of particles, so our extremization condition must be that $c_A+c_B$ is invariant under moving a particle from $z_1$ to $z_2$
\bea
0&=&\left(\frac{\delta}{\delta \rho(z_1)}-\frac{\delta}{\delta \rho(z_2)}\right)\left(c_A+c_B\right)=V(z_1)-V(z_2)\label{veq}\\
V(z)&=&\int_0^1 dx \left(\rho(x)  {\rm{ln}}\left|\s{\pi\alpha\left(x-z\right)}\right|-(1-\rho(x))  {\rm{ln}}\left|\s{\pi(2-\alpha)\left(x-z\right)}\right)\right|.\nonumber
\eea
Therefore one finds that extremization requires that the potential must be $z$-independent, or in other words that the force $F(z)$ vanishes
\bea
0&=&F(z)=\frac{\partial}{\partial z}V(z)\nonumber\\
&=&-\pi\int_0^1 dx \left[\rho(x)\left(\alpha{\rm{cot}}\left(\pi\alpha (x-z)\right)+(2-\alpha){\rm{cot}}\left(\pi(2-\alpha) (x-z)\right)\right)\right.\nonumber\\
&&+(\alpha-2){\rm{cot}}\left(\pi(2-\alpha) (x-z)\right)].\label{appeq}
\eea

We solve this equation for $\rho$ by discretizing, and reducing it to a linear matrix equation.  The equation is solved by inverting the operator which multiplies $\rho$ and multiplying it by the term on the last line.  The dimension of this matrix is independent of $N$, and so $N$ has disappeared from the problem.  We discretized into 25 parts, so that $\rho(x)$ is defined by at 25 positions.  Then we solved the equation to obtain our preferred value of $\rho(x)$, displayed in Fig.~\ref{rhofig}.  At various values of $N$, we then created a subset $F\in [1,N]$ with local density $\rho$.  

The results are reported in red in Fig.~\ref{2fig}.   As a result of this clumsy optimization procedure, our results are quite noisy.  In general at high $\alpha$ it systematically underestimates $\psi_0$.  However at low values of $\alpha$ it is closer to $\psi_0$ than $\psi_{\da\cdots\da}$.

\section{Conclusions}

State of the art computations of coordinate matrix elements in integrable models \cite{llmat,brock1} are only able to compute matrix elements with respect to coordinate states that are in some sense homogenous or linear, such as N\'eel states.   Our goal is to generalize these computations to states which are piecewise linear.   The strategy that we adopt is the decomposition of the lattice sites or particle positions into piecewise linear bins, which induces a decomposition of sum in the Coordinate  Bethe Ansatz.

More precisely, the CBA computes matrix elements as a sum of $N!$ terms, corresponding to maps
\beq
P:[1,N]\longrightarrow[1,N]
\eeq
or equivalently to elements of the symmetric group $S_N$.   Summing over all $N!$ terms in general is difficult.  But we found that if the domain $[1,N]$ is divided into bins $\ms_i$ and one fixes the images $P(\ms_i)$, the corresponding subset of terms can be summed exactly.  Each such subset is a coset $H\subset S_N$ and $S_N$ is the disjoint union of these cosets.  We provided this sum in Eq.~(\ref{master}).    Thus the task of summing over all $N!$ group elements is reduced to the task of summing over the cosets, of which there are only of order $2^N$.

More critically, in the case treated in Sec.~\ref{consez} in which the density was constant, the sums over all cosets had the same phase.  Thus one needed only to sum over the dominant cosets to obtain a good approximation.  More generally, in Sec.~\ref{doubsez} the sums over cosets had the same phase up to a sign.  Our approximations to the matrix elements were generally too high because we did not consider the fluctuation in signs.  

However for densities near the mean density, which are those that will contribute to wave functions in the large $N$ limit, the fluctuations are caused only by the terms near the boundaries of the image $[1,N]$.  We thus feel that it should be possible to develop an approximation scheme, along the lines of that developed in Sec.~\ref{consez} for transitions between $\ua$ and $\da$, which allows us to estimate the correction caused by these sign flips.  This will be done in future work.

Another critical generalization will be to other Hamiltonian eigenstates.  Our motivation, understanding mass gaps in quantum field theory, suggests that we only need to investigate low lying states.  For these, only a small number of momenta are changed and so we expect that we only need to modify our approximation scheme to keep track of which bin contains these excitations or, more simply, one may separately sum over $P^{-1}$ of the excited momenta.   The more challenging step comes next, we will try to generalize these results to the Lieb-Liniger model.   As the Tonks-Girardeau model is the strong coupling limit of the Lieb-Liniger model, we will be able to test our results. If this can be achieved, we expect that spin chains such as the Heisenberg XXX model should present no new conceptual difficulties, at least at $s=1/2$.

\section* {Acknowledgement}

\noindent
JE is supported by the CAS Key Research Program of Frontier Sciences grant QYZDY-SSW-SLH006 and the NSFC MianShang grants 11875296 and 11675223.  JE also thanks the Recruitment Program of High-end Foreign Experts for support.


\end{document}